\documentclass[fleqn,usenatbib]{mnras}

\usepackage{newtxtext,newtxmath}
\usepackage{adjustbox}
\usepackage[dvipsnames]{xcolor}

\usepackage[T1]{fontenc}

\DeclareRobustCommand{\VAN}[3]{#2}
\let\VANthebibliography\thebibliography
\def\thebibliography{\DeclareRobustCommand{\VAN}[3]{##3}\VANthebibliography}

\usepackage{graphicx}	
\usepackage{amsmath}	

\title[The BRITE view of CP1/2/3/4 stars]{Chemically peculiar stars investigated by the BRITE Mission}

\author[T. Begari et al.]{
Teja Begari,$^{1}$
Klaus Bernhard,$^{1}$
Ernst Paunzen$^{2}$\thanks{E-mail: epaunzen@physics.muni.cz (EP)}
and Prapti Mondal$^{2}$
\\
$^{1}$ American Association of Variable Star Observers, 185 Alewife Brook Parkway, Suite 410, Cambridge, MA 02138, USA\\
$^{2}$Department of Theoretical Physics and Astrophysics - Masaryk University, Kotlářská 2, Brno, Czechia\\
}

\date{Accepted 2026 January 24. Received 2026 January 08; in original form 2025 December 11}

\pubyear{\the\year{}}

\begin{document}
\label{firstpage}
\pagerange{\pageref{firstpage}--\pageref{lastpage}}
\maketitle

\begin{abstract}
We present a comprehensive analysis of BRITE photometry for 85 chemically peculiar stars, aimed at refining or determining their rotational periods. Utilizing a uniform Lomb-Scargle-based pipeline, we derived significant periods for 47 targets. A comparison with existing literature periods reveals generally good agreement, although several stars exhibit discrepant or previously unrecognized behavior. Notably, six targets display clear multiperiodicity, which, when combined with archival TESS data, suggests that these six candidates are likely misclassified, for example, as a magnetic CP2 or a CP4 star and instead exhibit characteristics consistent with a Be/shell star. Furthermore, eleven stars show no detectable periodic variations within the precision limits of BRITE. Our analysis demonstrates the effectiveness of long-term nanosatellite photometry, particularly when complemented by TESS data, in verifying catalogue periods, identifying multiperiodic behavior, and detecting potential misclassifications among bright CP stars.
\end{abstract}




\section{Introduction}

The upper main-sequence stars harbor a distinct population of chemically peculiar (CP) stars, constituting up to 15 percent of this stellar group. The anomalies in CP stars are primarily attributed to atomic diffusion processes in radiatively stable atmospheres \cite{michaud1970diffusion}, often modulated by strong magnetic fields that suppress convection and lead to stratified chemical distributions (\citealt{preston1974chemically}, \citealt{babel1992magnetically}). Characterized by anomalous spectral features indicative of unusual elemental abundance patterns, CP stars have been categorized into distinct subgroups:

\textit{CP1 (Metallic-line (Am) stars)} include A- to F-type stars whose spectra show weaker \ion{Ca}{II} K lines, and enhanced iron and heavier elements in their spectra than normal stars. As a result, the spectral types derived from the \ion{Ca}{II} K line is earlier by five or more spectral subclasses than those derived from the metallic-line spectrum \citep{2009ssc..book.....G}.
Many Am stars occur in close binary systems (\citealt{2005A&A...443..627T}, \citealt{2009AJ....138...28A}). 
Photometric variability among Am stars is generally weak or absent, as they lack strong magnetic fields and chemical spots. Some members, however, exhibit low-amplitude $\delta$~Scuti or $\gamma$~Doradus-type pulsations (\citealt{1989MNRAS.238.1077K}, \citealt{2011A&A...535A...3S}).

\textit{CP2 (Ap) stars} are late B- to early F-type objects showing overabundances of Si, Cr, Sr and Eu, often accompanied by strong, stable magnetic fields. Surface chemical spots cause periodic spectroscopic and photometric variations with the rotation period (\citealt{1947ApJ...105..283D}, \citealt{1950MNRAS.110..395S}). The rotational modulation typically results in strictly periodic light curves with amplitudes of a few hundredths up two a few tenth of a magnitude (\citealt{2018A&A...619A..98H}), occasionally superimposed by rapid oscillations (roAp pulsations). 

\textit{CP3 (HgMn) stars} display large overabundances of Hg and/or Mn, together with other iron-peak elements, but show no detectable large-scale magnetic fields. Their atmospheres appear comparatively stable, and they are often considered as hotter analogues of CP1 stars (\citealt{2018MNRAS.480.2953G}). Photometrically, HgMn stars are usually constant; nevertheless, subtle rotational modulations and possible pulsations have been reported from space-based photometry (\citealt{2018MNRAS.474.2467H}).

\textit{CP4 (He-weak/He-strong) stars} are mainly B-type stars that exhibit either a deficiency or an excess of helium in their spectra relative to normal stars of similar temperature. Like CP2 stars, they typically show strong magnetic fields that produce spectral and photometric variability linked to surface inhomogeneities \citep{1983ApJS...53..151B,1987ApJ...323..325B}. Their brightness changes arise from the rotation of chemically spotted stellar surfaces with uneven helium and metal distributions \citep{1990ApJ...365..665S}. Typical amplitudes are of the order of a few hundredths of a magnitude, with periods corresponding to the stellar rotation.

The BRITE (BRIght Target Explorer) Constellation is the first coordinated mission of nano-satellites intended for accurate photometric studies of the brightest stars. The mission was founded through international cooperation between Austria, Canada, and Poland, aiming to explore stellar variability and internal structure through long-term observation of stars commonly showing visual magnitudes above six. The constellation consists of five functional cubical satellites with dimensions of 20×20×20 cm, each housing a compact aperture telescope and a CCD camera optimized for red (transmitting in the range 550$-$700\,nm) or blue (390$-$460\,m) optical wavelengths. 
The BRITE constellation consists of three satellites with
a red filter, named BRITE-Toronto (BTr), UniBRITE (UBr), and
BRITE-Heweliusz (BHr), and two satellites equipped with a blue
filter, named BRITE-Austria (BAb) and BRITE-Lem (BLb).
In low Earth orbit, the BRITE satellites allow for high-cadence and uninterrupted photometric observations over long timescales \citep{2024A&A...683A..49Z}. This allows them to be especially efficient in the detection of low-amplitude oscillations and rotational modulation within both massive and evolved stellar populations. Ever since its foundation in 2013, the BRITE constellation has greatly contributed to the understanding of stellar pulsations, surface properties, rotational behavior, and magnetic activity of bright stars, successfully overcoming the issues faced by larger missions in dealing with brightness limitations \citep{2021Univ....7..199W}.

\begin{figure}
    \centering
    \includegraphics[width=1.0\linewidth]{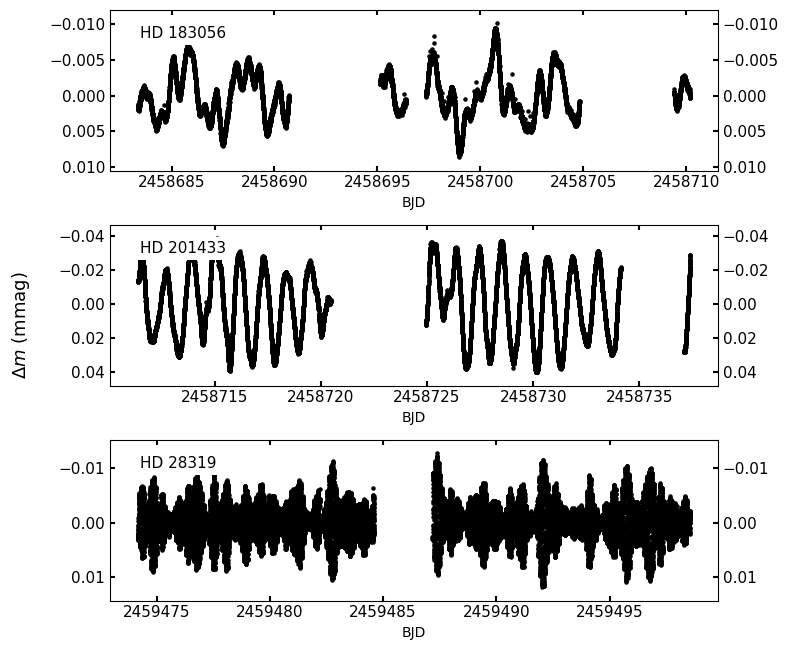}
    \caption{Example TESS Light curves of the three of the six multiperiodic stars.}
    \label{fig:TESS_multiperiodic-stars}
\end{figure}

We found six candidates to be multiperiodic, to better understand their multiperiodicity we used the TESS data. TESS (Transiting Exoplanet Survey Satellite) provides high-precision, nearly continuous space-based photometry across large portions of the sky and is therefore exceptionally well suited for studying periodic and multiperiodic brightness variations in a wide range of stellar classes \citep{2015ESS.....350301R}. Three example light curves of the six multiperiodic stars are presented in Fig.\ref{fig:TESS_multiperiodic-stars}. The mission observes the sky in a sector-by-sector strategy, delivering uniformly calibrated light curves with a stable instrumental response. For the present analysis, the corresponding TESS data products for suspected multiperiodic stars were retrieved from the Mikulski Archive for Space Telescopes (MAST), which offers standardized access to all publicly available TESS observations and associated metadata.

\begin{figure}
    \centering
    \includegraphics[width=1.0\linewidth]{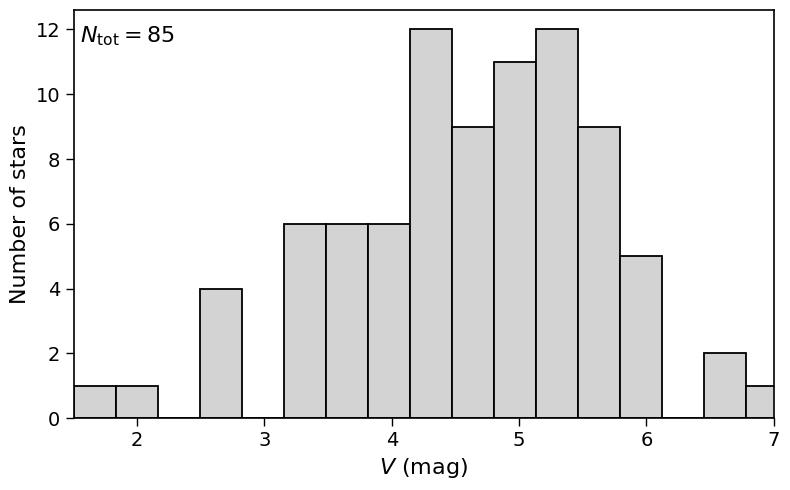}
    \caption{Histogram of the $V$ magnitudes for our target star sample.}
    \label{fig:Vmag}
\end{figure}

\section{data sources and target selection}
Our investigation into the CP stars leveraged the most recent and comprehensive compilation available, namely the General Catalogue of CP Stars by \cite{2009A&A...498..961R}. This catalogue provides a thorough and authoritative listing of chemically peculiar stars, allowing us to conduct a thorough search for CP1/2/3/4 stars (Table \ref{table_master1}). Given that the BRITE observations were focused on bright stars, see Fig.\ref{fig:Vmag}, which have been extensively studied and characterized, the classifications in this catalogue are likely to be robust and reliable, providing a solid foundation for our analysis.

In total, we used 466 individual datasets (listed in Table \ref{obs_log}) from all five satellites. They
accumulate to about 5.5 million data points.

\section{Time series analysis}
The Time series analysis was performed for all the candidates of interest with the program package PERANSO \cite{2016AN....337..239P}. The significant periods were identified by the Generalized Lomb–Scargle algorithm (\cite{2009A&A...496..577Z}). The Lomb-Scargle algorithm is a variant of the Discrete Fourier Transform (DFT) that transforms a finite sequence of unequally sampled data points from a function into a set of coefficients representing a linear combination of sine and cosine functions. This transformation maps the data from the time domain to the frequency domain, yielding a Lomb-Scargle periodogram that is time-shift invariant. From a statistical perspective, the periodogram can be interpreted as related to the $\chi^2$  statistic for a least-squares fit of a sinusoid to the data, which accommodates heteroscedastic measurement errors. Notably, the underlying frequency model is non-linear, and the basis functions at different frequencies do not exhibit orthogonality.

\subsection{Instrumental effects and data cleaning}

The BRITE light curves are notably affected by various celestial and terrestrial cycles, including the Sun, Moon, and daily cycles. These external influences can introduce periodic patterns and trends in the data. $\alpha^2$ Canum Venaticorum (ACV) variables, on the other hand, exhibit a remarkably wide range of periodic behavior, spanning from relatively short periods of around half a day (\cite{2017MNRAS.466.1399H}) to extremely long periods lasting hundreds or even thousands of years (\cite{2017A&A...601A..14M}). However, a review of existing literature suggests that most ACV variables typically display periods on the order of days (\cite{2017MNRAS.468.2745N}). Given the nature of these variables, the BRITE orbital period of 14.66 cycles per day (corresponding to a period of approximately 97-98 minutes) is unlikely to significantly impact our analysis of the longer-period ACV variables, as its effects will be most pronounced at higher frequencies.

To further explore this matter, we conducted a comprehensive analysis of all 85 light curves to identify the most prominent periods introduced by instrumental effects. We employed the Lomb-Scargle algorithm to search for signals within the 20 to 1000 day range. The data were then pre-whitened with the most significant periods detected, and the Lomb-Scargle algorithm was reapplied to uncover additional signals. This iterative process was repeated five times, allowing us to progressively uncover and remove the most influential instrumental periods from the data.
We applied a standardized, automated procedure to all light curves, consisting of the following steps:

(i) Initial cleaning: We used a basic $\sigma$-clipping algorithm to remove outliers from the raw light curve.

(ii) Pre-whitening: We removed the five most significant periods with the lowest False Alarm Probability (FAP) in the 50-1000 day range, as well as the one-day period, to mitigate dominant signals.

(iii) Secondary cleaning: We applied another $\sigma$-clipping algorithm to the residual light curve to further refine the data.

(iv) Period search: Finally, we searched for significant periods in the cleaned and pre-whitened data.

In addition, all data sets were analysed using a custom Python pipeline based on the generalized Lomb--Scargle periodogram implemented in \textsc{astropy} (\citealt{2018ApJS..236...16V}). 
After sorting the photometric data by time, an iterative $3\sigma$ clipping procedure based on the median absolute deviation was applied to remove outliers and spurious measurements. 
A uniformly sampled frequency grid between 0.01 and 3 cycles (\,c\,) d$^{-1}$ was computed with an oversampling factor of 50. For each data set, the two most significant peaks were identified, and their false-alarm probabilities were estimated.

These procedures allowed us to systematically process and analyse the light curves.

\subsection{Variable stars}
\begin{figure}
    \centering
    \includegraphics[width=1.0\linewidth]{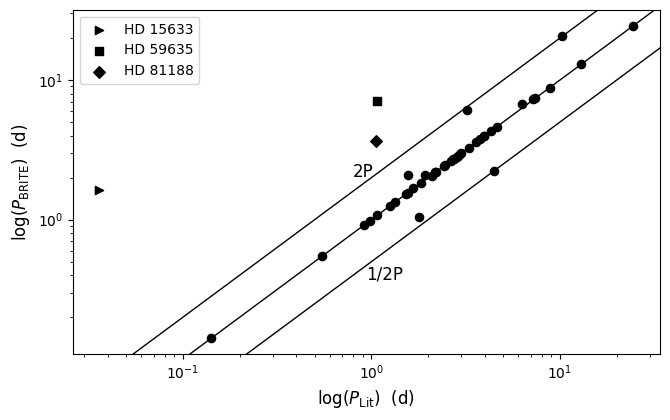}
    \caption{A comparison of BRITE-derived periods with literature values is presented for our sample stars. The plot includes lines indicating perfect agreement (unity line) as well as half- and double-period relationships. Three extreme outlier stars (HD 15633, HD 59635 and HD 81188), discussed in the text, are denoted by unique symbols.}
    \label{fig:LitVsBrite}
\end{figure}

\begin{figure}
    \centering
    \includegraphics[width=1.0\linewidth]{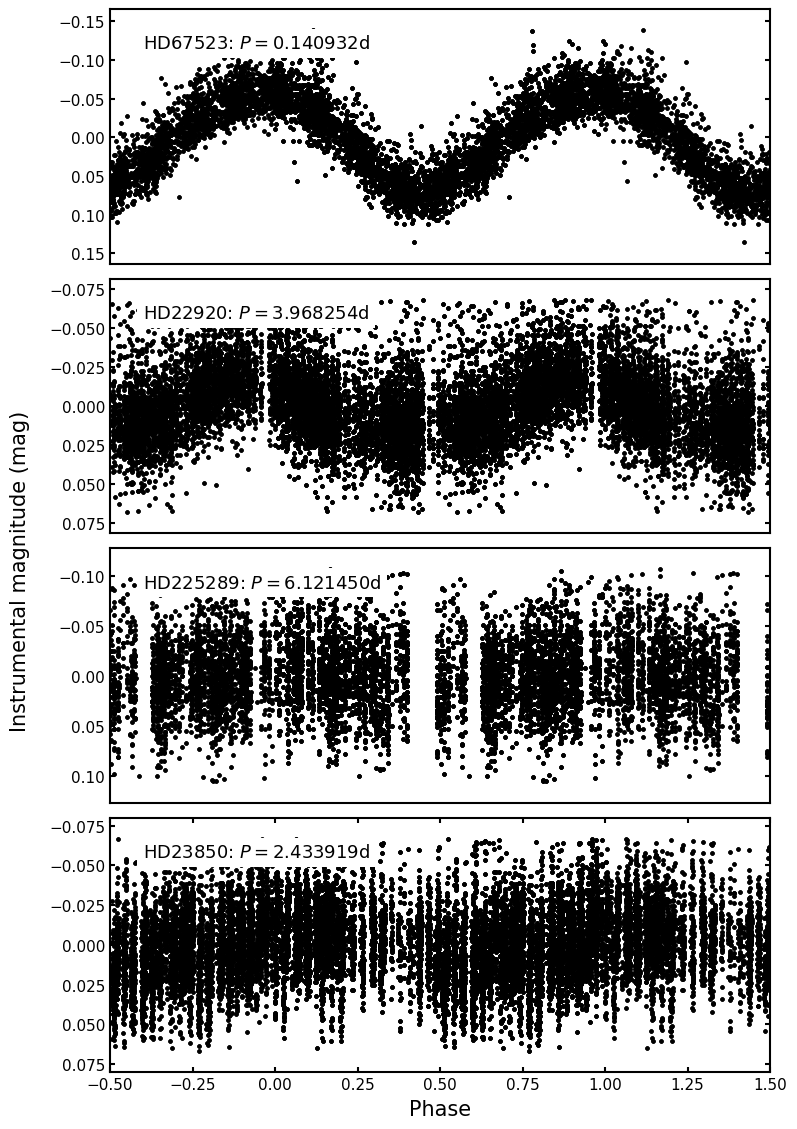}
    \caption{Phased light curves of four sample stars observed by BRITE, showcasing diverse periods and light-curve characteristics.}
    \label{fig:PhaseVsMag}
\end{figure}

In total we identified 74 variable stars, from which 51 have periods in the International Variable Star Index (VSX, \cite{2006SASS...25...47W}) of the American Association of Variable Star Observers. Out of 74 stars with periods in the literature, we collected the periods of the other 20 stars from \cite{2017MNRAS.468.2745N} and \cite{2023ApJS..268....4F}. For the candidate HD 128898, we collected the corrected rotational period from an extensive photometric and spectroscopic study by \cite{2024A&A...688A..62K}. For the candidate HD 40312, we collected the rotational period from \cite{2019MNRAS.483.3127S} and, for HD 53244, we collected the rotational period from \cite{2021MNRAS.506.5328K}. Example phased light curves of four sample stars (HD 67523, HD 22920, HD 225289, HD 23850) in Fig.\ref{fig:PhaseVsMag}. Each of them are CP1, CP2, CP3 and CP4 stars respectively. In Fig.\ref{fig:LitVsBrite}, we present a comparison of the here derived periods to the literature periods of the stars listed in Table \ref{tab_periods}. The typical period error amounts to 0.0001 d. Because in orientations where two spots come
into view during a star’s rotation cycle, a significant number of ACV variables exhibit double-waved light curves (\cite{Hümmerich_2016}). Therefore, a twice longer (or shorter) rotation period is sometimes
possible, in particular for objects with very small amplitudes and/or
significant scatter in their light curves.  We emphasize that the given
period values correspond to the periods with the highest FAP values; nevertheless, the most prominent periods may sometimes represent
harmonics of the true rotation period, as becomes obvious from
Fig.\ref{fig:LitVsBrite}. In ambiguous cases, the true period has to be determined by
other means such as radial velocity studies. 
There are only three obvious outliers in Fig. \ref{fig:LitVsBrite}, which are discussed in more detail below.

{\it HD 15633:} The TESS data clearly show $\delta$ Scuti type pulsation with several frequencies.
Although a long period similar to ours is also present. The star is reported as a CP1-type
object for which pulsation is a known phenomenon \citep{2013MNRAS.429..119P}. A detailed follow-up
analysis is needed to shed more light on this interesting star.

{\it HD 59635}: It shows the most extreme disagreement between the BRITE (7.063\,d) and literature \citep[0.9430 d;][]{2021MNRAS.504.3758P} periods. The BRITE light curve is characterized by a coherent single-wave modulation, entirely consistent with rotational variability in a magnetic CP star. The period of 7 d is robust and well supported by the BRITE frequency spectrum. The published period of 0.9430 d is almost certainly the result of ground-based daily aliasing, a well-documented issue for ACV stars whose periods lie close to integer multiples of the sidereal day. The alias cannot be reconciled with the BRITE signal or with any of its harmonics. We therefore conclude that the BRITE period represents the true stellar rotation period.

{\it HD 81188}: This star also exhibits a substantial difference between the BRITE (3.656 d) and the literature 
\citep[1.05729 d;][]{2023ApJS..268....4F} periods, which is a discrepancy of more than a factor of three. Inspection of the BRITE light curve reveals a well-phased, stable modulation consistent with rotational variability in magnetic Ap stars. The period difference may arise from early ground-based studies misinterpreting a harmonic of the true rotational frequency, a known issue for stars with low-amplitude or double-wave light curves. Unlike the misclassified multiperiodic objects discussed in Section 3.4, HD 81188 does not show additional frequencies in the BRITE data, further supporting its interpretation as a stable, monoperiodic ACV variable with a corrected period.





\begin{table*}
\centering
\caption{BRITE-derived periods ($P_{\rm BRITE}$) and literature periods ($P_{\rm Lit}$).}
\label{tab_periods}
\begin{tabular}{ccc|ccc|ccc}
\hline
\textbf{HD} & $P_{\rm BRITE}$ (d) & $P_{\rm Lit}$ (d) &
\textbf{HD} & $P_{\rm BRITE}$ (d) & $P_{\rm Lit}$ (d) &
\textbf{HD} & $P_{\rm BRITE}$ (d) & $P_{\rm Lit}$ (d) \\
\hline
\textbf{6961}   & --            & --         & \textbf{40183}  & 3.953   & 3.9600514 & \textbf{157792} & --           & --         \\
\textbf{11415}  & multiperiodic            & 1.16555    & \textbf{40292}  & --      & 0.490042  & \textbf{157919} & --           & 1.70713    \\
\textbf{15633}  & 1.618         & 0.036041   & \textbf{40312}  & 3.619   & 3.619       & \textbf{159876} & --           & 2.29       \\
\textbf{18296}  & 2.849         & 2.88422    & \textbf{53244}  & 6.871   & 6.214        & \textbf{162374} & --           & 14.5203781 \\
\textbf{18866}  & --            & 5.25922    & \textbf{54118}  & 3.275   & 3.27533   & \textbf{165040} & --           & 0.6733     \\
\textbf{20320}  & --            & --         & \textbf{55719}  & --      & --        & \textbf{168733} & 6.712        & 6.3        \\
\textbf{22634}  & 0.5463        & 0.546875   & \textbf{56022}  & 0.9133  & 0.9183    & \textbf{169467} & --           & --         \\
\textbf{22920}  & 3.968         & 3.9474     & \textbf{56455}  & 2.074   & 1.9347    & \textbf{173648} & 4.286        & 4.3        \\
\textbf{23408}  & 20.68         & 10.288     & \textbf{59635}  & 7.063   & 0.9430   & \textbf{174638} & 12.95        & 12.944     \\
\textbf{23850}  & 2.434         & 2.4266     & \textbf{64740}  & 1.331   & 1.33      & \textbf{175362} & 7.4          & 7.34       \\
\textbf{25823}  & 7.372         & 7.227424   & \textbf{67523}  & 0.1409  & 0.1408809 & \textbf{176723} & --           & 0.111914   \\
\textbf{26961}  & 1.529         & 1.52       & \textbf{73634}  & --      & 2.23416   & \textbf{182255} & 1.261        & 1.2622     \\
\textbf{27463}  & 2.798         & 2.83507    & \textbf{74560}  & 1.551   & 1.56235   & \textbf{182568} & 2.716/multiperiodic        & 2.7172     \\
\textbf{27628}  & 1.078         & 1.0718     & \textbf{81188}  & 3.656   & 1.05729   & \textbf{183056} & multiperiodic& 1.05171    \\
\textbf{27962}  & --            & 57.25      & \textbf{82434}  & 2.089   & 1.57095   & \textbf{189178} & --           & --         \\
\textbf{28319}  & multiperiodic & 0.07564    & \textbf{93030}  & 2.203   & 2.2026    & \textbf{189849} & --           & --         \\
\textbf{28527}  & --            & 1.278      & \textbf{104671} & 24.53   & 24.5  & \textbf{198639} & --           & 0.89601    \\
\textbf{29140}  & 3.571         & 3.5712     & \textbf{109026} & 2.726   & 2.72926   & \textbf{201433} & multiperiodic& 1.127      \\
\textbf{29305}  & 2.945         & 2.94247    & \textbf{120640} & 2.378   & 2.274600   & \textbf{201601} & --           & 0.00811    \\
\textbf{29388}  & --            & --         & \textbf{122532} & 1.834   & 1.837     & \textbf{202444} & --           & 1.19213    \\
\textbf{30780}  & --            & 0.042      & \textbf{125823} & 8.783   & 8.8171    & \textbf{204188} & --           & 0.044      \\
\textbf{32549}  & 4.64          & 4.6397     & \textbf{128898} & 2.24    & 4.479   & \textbf{205924} & --           & --         \\
\textbf{33641}  & --            & 0.047805   & \textbf{135379} & --      & 3.17411   & \textbf{206155} & 2.622        & 2.62821423 \\
\textbf{33904}  & 3.011         & 2.9895     & \textbf{138769} & 2.065   & 2.0894    & \textbf{209790} & --           & 4.9125     \\
\textbf{33959}  & --            & 0.088088   & \textbf{141556} & 1.04    & 1.79292   & \textbf{211336} & --           & 0.078524   \\
\textbf{34452}  & 2.466         & 2.466      & \textbf{142990} & 0.9789  & 0.9789    & \textbf{223128} & --           & 0.12959    \\
\textbf{35039}  & multiperiodic            & 41.033     & \textbf{150549} & 3.771   & 3.76      & \textbf{225289} & 6.121        & 3.22072    \\
\textbf{35497}  & --            & --         & \textbf{152564} & 2.201   & 2.1637    & \textbf{ }     &              &            \\
\textbf{36960}  & --            & --         & \textbf{155203} & --      & 0.55682   & \textbf{ }     &              &            \\
\hline
\end{tabular}
\end{table*}

\subsection{Apparently constant stars}
A subset of eleven objects exhibited no significant periodic signals in either the BRITE data or in the literature. The amplitude spectra of these stars were consistent with noise within the instrumental precision of BRITE, and no peaks with a false-alarm probability below the standard significance threshold were detected in the cleaned and pre-whitened light curves. While the absence of detectable signals is consistent with expectations for CP1 and CP3 stars, which are typically non-variable or exhibit low photometric amplitudes, the lack of variability in suspected CP2 and CP4 stars warrants caution regarding their classifications, particularly in the absence of robust spectroscopic confirmation. 
Another possibility is that we face super-slowly rotating Ap (ssrAp) stars \citep{2024A&A...692A.231H,2025A&A...703A.102M}. These are CP stars with rotational periods longer than 50 days. Therefore, they are difficult to detect. An example is HD 55719, a strongly magnetic CP2 star whose rotation period exceeds 38 yr \citep{2022MNRAS.514.3485G}.
Further high-resolution spectroscopic observations are necessary to accurately assess their chemical peculiarities and magnetic properties. In addition, much longer photometric time series are needed to detect
ssrAps, for example.

The noise in the amplitude spectra can be well described as 1/f or ''flicker'' or ''pink'' noise \citep{chatfield2004timeseries}. It can be described as a linear law by plotting the logarithm of the frequency ($f$) versus the logarithm of the amplitude ($A$).

\begin{equation}
\log A = a + b \log f.
\end{equation}

The $a$ and $b$ values of the eleven amplitude spectra without any significant peaks are listed in Table \ref{result_constant}. The slopes for the individual stars and filters differ significantly. This
means that the noise characteristics are dependent on the individual data sets. The level of the noise characterised by the intercept depends on the number of data points and the magnitude of the stars (photon statistics).

\begin{table*}
\begin{center}
\caption{Results from fitting the amplitude spectra of the apparently constant stars in the form $\log A = a + b \log f$, where $A$ is the amplitude and $f$ the frequency.}
\label{result_constant}
\begin{tabular}{c|cccc|cccc}
  \hline
  & \multicolumn{4}{c|}{Blue region} & \multicolumn{4}{c|}{Red region} \\
HD & $a$ & $\sigma(a)$ & $b$ & $\sigma(b)$ & $a$ & $\sigma(a)$ & $b$ & $\sigma(b)$ \\
\hline
\textbf{6961}	&	$-$3.382	&	0.002	&	$-$0.066	&	0.004	&	$-$3.238	&	0.003	&	$-$0.165	&	0.008	\\
\textbf{20320}	&		&		&		&		&	$-$3.584	&	0.007	&	$-$0.121	&	0.018	\\
\textbf{29388}	&	$-$3.065	&	0.006	&	$-$0.581	&	0.013	&		&		&		&		\\
\textbf{35497}	&	$-$3.150	&	0.007	&	$-$0.758	&	0.016	&	$-$2.370	&	0.049	&	$-$0.384	&	0.125	\\
\textbf{36960}	&	$-$2.431	&	0.006	&	$-$0.122	&	0.015	&	$-$2.070	&	0.006	&	$-$0.480	&	0.013	\\
\textbf{55719}	&		&		&		&		&	$-$3.768	&	0.005	&	$-$0.257	&	0.012	\\
\textbf{157792}	&	$-$2.847	&	0.001	&	$-$0.223	&	0.003	&	$-$3.180	&	0.001	&	$-$0.275	&	0.003	\\
\textbf{169467}	&	$-$2.620	&	0.001	&	$-$0.306	&	0.003	&	$-$2.755	&	0.001	&	$-$0.799	&	0.003	\\
\textbf{189178}	&	$-$3.276	&	0.005	&	$-$0.421	&	0.012	&	$-$3.622	&	0.001	&	$-$0.431	&	0.003	\\
\textbf{189849}	&	$-$2.632	&	0.006	&	$-$0.087	&	0.016	&	$-$3.538	&	0.002	&	$-$0.055	&	0.005	\\
\textbf{205924}	&		&		&		&		&	$-$3.545	&	0.006	&	$-$0.220	&	0.014	\\
\hline
\end{tabular}
\end{center}
\end{table*}

\subsection{Notes on misclassified CP2 and CP4 candidates}

Six targets listed as CP2 and CP4 stars in the General Catalogue of CP Stars show clear multiperiodic variability in our BRITE and/or corresponding TESS data. Since genuine CP2 and CP4 stars are expected to exhibit strictly monoperiodic light variations caused by rotational modulation of chemically spotted surfaces, the observed behaviour raises the possibility of misclassification.
A representative example is HD~11415 (CP4). Although included in the General Catalogue of CP Stars as a well-known B3 He-weak star, it is listed in SIMBAD as a Be/shell object with spectral type B3,Vp sh. This classification ultimately traces back to the study of  \citet{1968ApJS...17..371L}, which already noted spectral features incompatible with helium-weak or helium-strong objects.
Another instructive case is HD~182568 (2~Cyg). It shows a clear periodic signal in the BRITE data at the published literature value (type CP4), yet its light curve also reveals additional intrinsic variability. Such behaviour is inconsistent with the strictly monoperiodic rotational modulation expected for magnetic CP stars. A recent detailed spectroscopic analysis \citep{2025AJ....169...48C} demonstrated that HD~182568 is in fact a young B3,IV runaway binary system with no evidence of chemical peculiarity. Consequently, its classification in earlier CP catalogues is almost certainly incorrect.
Similar arguments apply to the remaining four candidates—HD~35039 (CP4), HD~28319 (CP2), HD~183056 (CP2), and HD~201433 (CP2). As a summary, we conclude that the evidence strongly suggests that all six objects are more plausibly SPB or Be/shell stars rather than genuine helium-peculiar CP2 or CP4 stars.

Another interesting possibility is the presence of a pulsating component in a binary system in
which the brighter component is a magnetic CP star. \citet{2020CoSka..50..570P} discussed the binary fraction of magnetic CP stars, concluding that up to 50\% of all magnetic CP stars are part of a binary system. However, there are only a handful of such possible systems reported \citep{2019MNRAS.487.4230S,2025arXiv250908426S}. To probe our six targets for binarity, additional radial velocity
and high-resolution spectroscopic observations are needed.

\section{Astrophysical Parameters and the Hertzsprung-Russell Diagram} \label{section_astro}

We here follow the approach presented by \citet{2021MNRAS.504.3758P}, which we will briefly review.

{\it Photometry:} photometric data of the Johnson $UBV$ and Geneva 7-colour systems were taken from
the All-Sky Compiled Catalogue of 2.5 million stars \citep[ASCC,][]{2001KFNT...17..409K} and 
\citet{2022A&A...661A..89P}, whereas the Str{\"o}mgren-Crawford $uvby\beta$ indices were extracted from the catalogue by \citet{2015A&A...580A..23P}.
 
{\it Reddening:} we used the commonly employed dereddening procedures published by \citet{1993A&A...268..653N}.

For the calibrations of the different photometric systems, we used the following relations \citep{2006A&A...458..293P}:
\begin{equation}
A_V = 3.1E(B-V) = 4.3E(b-y) = 4.95E(B2-V1).
\end{equation}

{\it Bolometric correction:} we used the correlation given by \citet{2008A&A...491..545N}, which was tailored for the use with CP stars.

{\it Luminosity:} the parallaxes available from both the {\it Hipparcos} \citep{2007A&A...474..653V} and the {\it Gaia} \citep{2018A&A...616A...1G,2023A&A...674A...1G} space missions were used. 
Because our sample consists of bright stars, the {\it Gaia} parallaxes have to be corrected.

{\it Effective temperature:} if available, data from the Johnson $UBV$, Geneva 7-colour, and Str{\"o}mgren-Crawford $uvby\beta$ photometric systems were used. \citet{2008A&A...491..545N} introduced calibrations for CP stars using individual corrections for the temperature domain and the CP subclass, which are summarised in their Table 2. We here follow their approach. For the derivation of the final effective temperatures, all calibrated values were averaged and the standard deviations were calculated.

\begin{figure}
    \centering
    \includegraphics[width=1.0\linewidth]{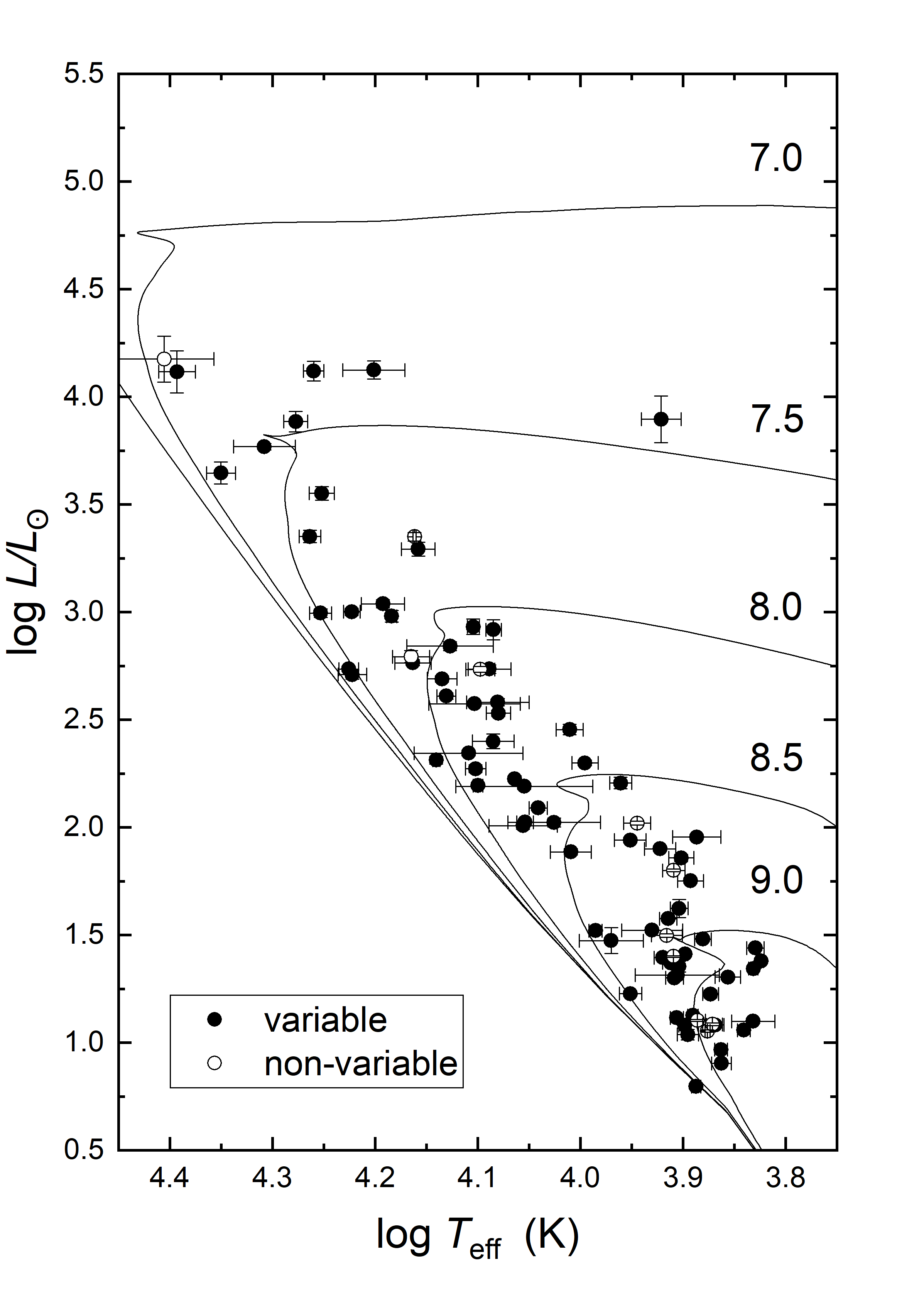}
    \caption{Positions of our target stars in the Hertzsprung-Russell diagram, which is provided in the form of $\log T_\mathrm{eff}$ versus $\log L/L_\odot$ (Table \ref{table_master1}). Also shown are stellar evolutionary models for the indicated logarithmic ages by \citet{2012MNRAS.427..127B}.}
    \label{fig:hrd_BRITE-CP}
\end{figure}

In Fig. \ref{fig:hrd_BRITE-CP}, the Hertzsprung-Russell diagram is presented together with the isochrones from \citet{2012MNRAS.427..127B} for [Z]\,=\,0.0152. As can be seen, only a few objects are very close to the 
zero-age main-sequence. A fact which was found before \citep{2020A&A...640A..40H}.

\section{Conclusions}

We conducted a comprehensive analysis of BRITE photometry for a sample of 85 chemically peculiar stars across the CP1, CP2, CP3, and CP4 classes. Employing a uniform time-series analysis approach, which included iterative cleaning, pre-whitening, and Lomb-Scargle analysis, we derived reliable rotational periods for 47 stars. Our analysis revealed six targets exhibiting clear multiperiodic behavior, which was further confirmed by complementary TESS photometry. This behavior is inconsistent with the expected monoperiodic rotational modulation of magnetic CP stars, suggesting that six of these stars (HD 11415, HD 35039, HD 182568, HD 28319, HD 183056 and HD 201433) are likely misclassified Be/shell stars. This finding highlights the importance of revisiting historical classifications with modern space-based data. Additionally, ten stars showed no significant photometric variability within the precision limits of BRITE. While this is consistent with expectations for CP1 and CP3 stars, the absence of periodic modulation in some CP2/CP4 stars may indicate that their classifications require spectroscopic revision. The sample's placement in the Hertzsprung-Russell diagram confirms that most variable stars occupy the expected region of the main sequence for magnetic and chemically peculiar objects, demonstrating the effectiveness of BRITE photometry in detecting low-amplitude rotational variability in bright CP stars. Our results underscore the value of combining long-term nanosatellite monitoring with complementary TESS data to refine classifications, identify misclassified objects, and advance our understanding of the surface structures and rotational properties of chemically peculiar stars.

\section*{Acknowledgements}

The authors thank Gautier Mathys for his valuable comments, which helped improve the quality of this paper. We also thank Stefan Hümmerich for his helpful comments on some aspects of this paper. This paper includes data collected with the TESS mission, obtained from the MAST data archive at the Space Telescope Science Institute (STScI). Funding for the TESS mission is provided by the NASA Explorer Program. STScI is operated by the Association of Universities for Research in Astronomy, Inc., under NASA contract NAS 5–26555.

\section*{Data Availability}

The data underlying this article will be shared on reasonable request
to the corresponding author.



\bibliographystyle{mnras}
\bibliography{example} 

@article{babel1992magnetically,
  title={Magnetically confined wind on the Ap star 53 Camelopardalis?},
  author={Babel, Jacques},
  journal={Astronomy and Astrophysics (ISSN 0004-6361), vol. 258, no. 2, p. 449-463.},
  volume={258},
  pages={449--463},
  year={1992}
}

@article{Hümmerich_2016,
doi = {10.3847/0004-6256/152/4/104},
url = {https://doi.org/10.3847/0004-6256/152/4/104},
year = {2016},
month = {oct},
publisher = {The American Astronomical Society},
volume = {152},
number = {4},
pages = {104},
author = {Hümmerich, Stefan and Paunzen, Ernst and Bernhard, Klaus},
title = {NEW PHOTOMETRICALLY VARIABLE MAGNETIC CHEMICALLY PECULIAR STARS IN THE ASAS-3 ARCHIVE},
journal = {The Astronomical Journal},

}

@ARTICLE{2013MNRAS.429..119P,
       author = {{Paunzen}, E. and {Wraight}, K.~T. and {Fossati}, L. and {Netopil}, M. and {White}, G.~J. and {Bewsher}, D.},
        title = "{A photometric study of chemically peculiar stars with the STEREO satellites - II. Non-magnetic chemically peculiar stars}",
      journal = {\mnras},
         year = 2013,
        month = feb,
       volume = {429},
       number = {1},
        pages = {119-125},
          doi = {10.1093/mnras/sts318},
archivePrefix = {arXiv},
       eprint = {1211.1535},
 primaryClass = {astro-ph.SR},
       adsurl = {https://ui.adsabs.harvard.edu/abs/2013MNRAS.429..119P}
}

@phdthesis{michaud1970diffusion,
  title={I. Diffusion processes in A-PEC stars. II. Nucleosynthesis in Si burning},
  author={Michaud, Georges Joseph},
  year={1970},
  school={California Institute of Technology}
}

@article{preston1974chemically,
  title={The chemically peculiar stars of the upper main sequence},
  author={Preston, George W},
  journal={In: Annual review of astronomy and astrophysics. Volume 12.(A75-13476 03-90) Palo Alto, Calif., Annual Reviews, Inc., 1974, p. 257-277.},
  volume={12},
  pages={257--277},
  year={1974}
}

@ARTICLE{2009A&A...498..961R,
       author = {{Renson}, P. and {Manfroid}, J.},
        title = "{Catalogue of Ap, HgMn and Am stars}",
      journal = {\aap},
         year = 2009,
        month = may,
       volume = {498},
       number = {3},
        pages = {961-966},
          doi = {10.1051/0004-6361/200810788},
       adsurl = {https://ui.adsabs.harvard.edu/abs/2009A&A...498..961R}
}

@ARTICLE{2016AN....337..239P,
       author = {{Paunzen}, E. and {Vanmunster}, T.},
        title = "{Peranso - Light curve and period analysis software}",
      journal = {Astronomische Nachrichten},
         year = 2016,
        month = mar,
       volume = {337},
       number = {3},
        pages = {239},
          doi = {10.1002/asna.201512254},
archivePrefix = {arXiv},
       eprint = {1602.05329},
 primaryClass = {astro-ph.IM},
       adsurl = {https://ui.adsabs.harvard.edu/abs/2016AN....337..239P}
}

@ARTICLE{2009A&A...496..577Z,
       author = {{Zechmeister}, M. and {K{\"u}rster}, M.},
        title = "{The generalised Lomb-Scargle periodogram. A new formalism for the floating-mean and Keplerian periodograms}",
      journal = {\aap},
         year = 2009,
        month = mar,
       volume = {496},
       number = {2},
        pages = {577-584},
          doi = {10.1051/0004-6361:200811296},
archivePrefix = {arXiv},
       eprint = {0901.2573},
 primaryClass = {astro-ph.IM},
       adsurl = {https://ui.adsabs.harvard.edu/abs/2009A&A...496..577Z}
}

@ARTICLE{2017MNRAS.466.1399H,
       author = {{H{\"u}mmerich}, Stefan and {Bernhard}, Klaus and {Paunzen}, Ernst and {Hambsch}, Franz-Josef and {Bohlsen}, Terry and {Powles}, Jonathan},
        title = "{An investigation of four chemically peculiar stars with photometric periods below 12 h}",
      journal = {\mnras},
         year = 2017,
        month = apr,
       volume = {466},
       number = {2},
        pages = {1399-1411},
          doi = {10.1093/mnras/stw3186},
archivePrefix = {arXiv},
       eprint = {1612.04708},
 primaryClass = {astro-ph.SR},
       adsurl = {https://ui.adsabs.harvard.edu/abs/2017MNRAS.466.1399H}
}

@ARTICLE{2017A&A...601A..14M,
       author = {{Mathys}, G.},
        title = "{Ap stars with resolved magnetically split lines: Magnetic field determinations from Stokes I and V spectra{\ensuremath{\star}}}",
      journal = {\aap},
         year = 2017,
        month = may,
       volume = {601},
          eid = {A14},
        pages = {A14},
          doi = {10.1051/0004-6361/201628429},
archivePrefix = {arXiv},
       eprint = {1612.03632},
 primaryClass = {astro-ph.SR},
       adsurl = {https://ui.adsabs.harvard.edu/abs/2017A&A...601A..14M}
}

@ARTICLE{2017MNRAS.468.2745N,
       author = {{Netopil}, Martin and {Paunzen}, Ernst and {H{\"u}mmerich}, Stefan and {Bernhard}, Klaus},
        title = "{An investigation of the rotational properties of magnetic chemically peculiar stars}",
      journal = {\mnras},
         year = 2017,
        month = jul,
       volume = {468},
       number = {3},
        pages = {2745-2756},
          doi = {10.1093/mnras/stx674},
archivePrefix = {arXiv},
       eprint = {1703.05218},
 primaryClass = {astro-ph.SR},
       adsurl = {https://ui.adsabs.harvard.edu/abs/2017MNRAS.468.2745N}
}

@ARTICLE{2005A&A...443..627T,
       author = {{Th{\'e}ado}, S. and {Vauclair}, S. and {Cunha}, M.~S.},
        title = "{Helium settling and mass loss in magnetic Ap stars. I. The chemical stratification}",
      journal = {\aap},
         year = 2005,
        month = nov,
       volume = {443},
       number = {2},
        pages = {627-641},
          doi = {10.1051/0004-6361:20052933},
       adsurl = {https://ui.adsabs.harvard.edu/abs/2005A&A...443..627T}
}

@ARTICLE{2009AJ....138...28A,
       author = {{Abt}, Helmut A.},
        title = "{Why are There Normal Slow Rotators Among A-Type Stars?}",
      journal = {\aj},
         year = 2009,
        month = jul,
       volume = {138},
       number = {1},
        pages = {28-32},
          doi = {10.1088/0004-6256/138/1/28},
       adsurl = {https://ui.adsabs.harvard.edu/abs/2009AJ....138...28A}
}

@ARTICLE{1989MNRAS.238.1077K,
       author = {{Kurtz}, D.~W.},
        title = "{Metallicism and pulsation : the discovery of delta Scuti variability in a classical AM star, HD 1097.}",
      journal = {\mnras},
         year = 1989,
        month = jun,
       volume = {238},
        pages = {1077-1084},
          doi = {10.1093/mnras/238.3.1077},
       adsurl = {https://ui.adsabs.harvard.edu/abs/1989MNRAS.238.1077K}
}

@ARTICLE{2011A&A...535A...3S,
       author = {{Smalley}, B. and {Kurtz}, D.~W. and {Smith}, A.~M.~S. and {Fossati}, L. and {Anderson}, D.~R. and {Barros}, S.~C.~C. and {Butters}, O.~W. and {Collier Cameron}, A. and {Christian}, D.~J. and {Enoch}, B. and {Faedi}, F. and {Haswell}, C.~A. and {Hellier}, C. and {Holmes}, S. and {Horne}, K. and {Kane}, S.~R. and {Lister}, T.~A. and {Maxted}, P.~F.~L. and {Norton}, A.~J. and {Parley}, N. and {Pollacco}, D. and {Simpson}, E.~K. and {Skillen}, I. and {Southworth}, J. and {Street}, R.~A. and {West}, R.~G. and {Wheatley}, P.~J. and {Wood}, P.~L.},
        title = "{SuperWASP observations of pulsating Am stars}",
      journal = {\aap},
         year = 2011,
        month = nov,
       volume = {535},
          eid = {A3},
        pages = {A3},
          doi = {10.1051/0004-6361/201117230},
archivePrefix = {arXiv},
       eprint = {1107.0246},
 primaryClass = {astro-ph.SR},
       adsurl = {https://ui.adsabs.harvard.edu/abs/2011A&A...535A...3S}
}

@ARTICLE{1947ApJ...105..283D,
       author = {{Deutsch}, Armin J.},
        title = "{A Study of the Spectrum Variables of Type a.}",
      journal = {\apj},
         year = 1947,
        month = mar,
       volume = {105},
        pages = {283},
          doi = {10.1086/144904},
       adsurl = {https://ui.adsabs.harvard.edu/abs/1947ApJ...105..283D}
}

@ARTICLE{1950MNRAS.110..395S,
       author = {{Stibbs}, D.~W.~N.},
        title = "{A study of the spectrum and magnetic variable star HD 125248}",
      journal = {\mnras},
         year = 1950,
        month = jan,
       volume = {110},
        pages = {395},
          doi = {10.1093/mnras/110.4.395},
       adsurl = {https://ui.adsabs.harvard.edu/abs/1950MNRAS.110..395S}
}

@ARTICLE{2018A&A...619A..98H,
       author = {{H{\"u}mmerich}, S. and {Mikul{\'a}{\v{s}}ek}, Z. and {Paunzen}, E. and {Bernhard}, K. and {Jan{\'\i}k}, J. and {Yakunin}, I.~A. and {Pribulla}, T. and {Va{\v{n}}ko}, M. and {Mat{\v{e}}chov{\'a}}, L.},
        title = "{The Kepler view of magnetic chemically peculiar stars}",
      journal = {\aap},
         year = 2018,
        month = nov,
       volume = {619},
          eid = {A98},
        pages = {A98},
          doi = {10.1051/0004-6361/201832938},
archivePrefix = {arXiv},
       eprint = {1808.05669},
 primaryClass = {astro-ph.SR},
       adsurl = {https://ui.adsabs.harvard.edu/abs/2018A&A...619A..98H}
}

@ARTICLE{2018MNRAS.480.2953G,
       author = {{Ghazaryan}, S. and {Alecian}, G. and {Hakobyan}, A.~A.},
        title = "{New catalogue of chemically peculiar stars, and statistical analysis}",
      journal = {\mnras},
         year = 2018,
        month = nov,
       volume = {480},
       number = {3},
        pages = {2953-2962},
          doi = {10.1093/mnras/sty1912},
archivePrefix = {arXiv},
       eprint = {1807.06902},
 primaryClass = {astro-ph.SR},
       adsurl = {https://ui.adsabs.harvard.edu/abs/2018MNRAS.480.2953G}
}

@ARTICLE{2022A&A...661A..89P,
       author = {{Paunzen}, E.},
        title = "{Catalogue of stars measured in the Geneva seven-colour photometric system}",
      journal = {\aap},
         year = 2022,
        month = may,
       volume = {661},
          eid = {A89},
        pages = {A89},
          doi = {10.1051/0004-6361/202142355},
archivePrefix = {arXiv},
       eprint = {2111.04810},
 primaryClass = {astro-ph.SR},
       adsurl = {https://ui.adsabs.harvard.edu/abs/2022A&A...661A..89P}
}

@ARTICLE{2015A&A...580A..23P,
   author = {{Paunzen}, E.},
    title = "{A new catalogue of Str{\"o}mgren-Crawford uvby{$\beta$} photometry}",
  journal = {\aap},
archivePrefix = "arXiv",
   eprint = {1506.04568},
 primaryClass = "astro-ph.SR",
     year = 2015,
    month = aug,
   volume = 580,
      eid = {A23},
    pages = {A23},
      doi = {10.1051/0004-6361/201526413},
   adsurl = {http://esoads.eso.org/abs/2015A%26A...580A..23P}
}

@ARTICLE{2001KFNT...17..409K,
   author = {{Kharchenko}, N.~V.},
    title = "{All-sky compiled catalogue of 2.5 million stars}",
  journal = {Kinematika i Fizika Nebesnykh Tel},
     year = 2001,
    month = oct,
   volume = 17,
    pages = {409-423},
   adsurl = {http://adsabs.harvard.edu/abs/2001KFNT...17..409K}
}

@ARTICLE{1993A&A...268..653N,
       author = {{Napiwotzki}, R. and {Schoenberner}, D. and {Wenske}, V.},
        title = "{On the determination of effective temperature and surface gravity of B, A, and F stars using Stromgren uvby-beta photometry.}",
      journal = {\aap},
         year = 1993,
        month = feb,
       volume = {268},
        pages = {653-666},
       adsurl = {https://ui.adsabs.harvard.edu/abs/1993A&A...268..653N}
}

@ARTICLE{2006A&A...458..293P,
       author = {{Paunzen}, E. and {Schnell}, A. and {Maitzen}, H.~M.},
        title = "{An empirical temperature calibration for the {\ensuremath{\Delta}} a photometric system. II. The A-type and mid F-type stars}",
      journal = {\aap},
         year = 2006,
        month = oct,
       volume = {458},
       number = {1},
        pages = {293-296},
          doi = {10.1051/0004-6361:20064889},
archivePrefix = {arXiv},
       eprint = {astro-ph/0607567},
 primaryClass = {astro-ph},
       adsurl = {https://ui.adsabs.harvard.edu/abs/2006A&A...458..293P}
}

@ARTICLE{2008A&A...491..545N,
       author = {{Netopil}, M. and {Paunzen}, E. and {Maitzen}, H.~M. and {North}, P. and
         {Hubrig}, S.},
        title = "{Chemically peculiar stars and their temperature calibration}",
      journal = {\aap},
         year = 2008,
        month = nov,
       volume = {491},
       number = {2},
        pages = {545-554},
          doi = {10.1051/0004-6361:200810325},
archivePrefix = {arXiv},
       eprint = {0809.5131},
 primaryClass = {astro-ph},
       adsurl = {https://ui.adsabs.harvard.edu/abs/2008A&A...491..545N}
}

@ARTICLE{2012MNRAS.427..127B,
       author = {{Bressan}, Alessandro and {Marigo}, Paola and {Girardi}, L{\'e}o. and
         {Salasnich}, Bernardo and {Dal Cero}, Claudia and {Rubele}, Stefano and
         {Nanni}, Ambra},
        title = "{PARSEC: stellar tracks and isochrones with the PAdova and TRieste Stellar Evolution Code}",
      journal = {\mnras},
         year = 2012,
        month = nov,
       volume = {427},
       number = {1},
        pages = {127-145},
          doi = {10.1111/j.1365-2966.2012.21948.x},
archivePrefix = {arXiv},
       eprint = {1208.4498},
 primaryClass = {astro-ph.SR},
       adsurl = {https://ui.adsabs.harvard.edu/abs/2012MNRAS.427..127B}
}

@ARTICLE{2021MNRAS.504.3758P,
       author = {{Paunzen}, E. and {Sup{\'\i}kov{\'a}}, J. and {Bernhard}, K. and {H{\"u}mmerich}, S. and {Pri{\v{s}}egen}, M.},
        title = "{Magnetic chemically peculiar stars investigated by the Solar Mass Ejection Imager}",
      journal = {\mnras},
         year = 2021,
        month = jul,
       volume = {504},
       number = {3},
        pages = {3758-3772},
          doi = {10.1093/mnras/stab1100},
archivePrefix = {arXiv},
       eprint = {2105.02206},
 primaryClass = {astro-ph.SR},
       adsurl = {https://ui.adsabs.harvard.edu/abs/2021MNRAS.504.3758P}
}

@ARTICLE{2018A&A...616A...1G,
       author = {{Gaia Collaboration} and {Brown}, A.~G.~A. and {Vallenari}, A. and
         {Prusti}, T. and {de Bruijne}, J.~H.~J. and {Babusiaux}, C. and
         {Bailer-Jones}, C.~A.~L. and {Biermann}, M. and {Evans}, D.~W. and
         {Eyer}, L. and {Jansen}, F. and {Jordi}, C. and {Klioner}, S.~A. and
         {Lammers}, U. and {Lindegren}, L. and {Luri}, X. and {Mignard}, F. and
         {Panem}, C. and {Pourbaix}, D. and {Randich}, S. and {Sartoretti}, P. and
         {Siddiqui}, H.~I. and {Soubiran}, C. and {van Leeuwen}, F. and
         {Walton}, N.~A. and {Arenou}, F. and {Bastian}, U. and {Cropper}, M. and
         {Drimmel}, R. and {Katz}, D. and {Lattanzi}, M.~G. and {Bakker}, J. and
         {Cacciari}, C. and {Casta{\~n}eda}, J. and {Chaoul}, L. and
         {Cheek}, N. and {De Angeli}, F. and {Fabricius}, C. and {Guerra}, R. and
         {Holl}, B. and {Masana}, E. and {Messineo}, R. and {Mowlavi}, N. and
         {Nienartowicz}, K. and {Panuzzo}, P. and {Portell}, J. and
         {Riello}, M. and {Seabroke}, G.~M. and {Tanga}, P. and
         {Th{\'e}venin}, F. and {Gracia-Abril}, G. and {Comoretto}, G. and
         {Garcia-Reinaldos}, M. and {Teyssier}, D. and {Altmann}, M. and
         {Andrae}, R. and {Audard}, M. and {Bellas-Velidis}, I. and
         {Benson}, K. and {Berthier}, J. and {Blomme}, R. and {Burgess}, P. and
         {Busso}, G. and {Carry}, B. and {Cellino}, A. and {Clementini}, G. and
         {Clotet}, M. and {Creevey}, O. and {Davidson}, M. and {De Ridder}, J. and
         {Delchambre}, L. and {Dell'Oro}, A. and {Ducourant}, C. and
         {Fern{\'a}ndez-Hern{\'a}ndez}, J. and {Fouesneau}, M. and
         {Fr{\'e}mat}, Y. and {Galluccio}, L. and {Garc{\'\i}a-Torres}, M. and
         {Gonz{\'a}lez-N{\'u}{\~n}ez}, J. and {Gonz{\'a}lez-Vidal}, J.~J. and
         {Gosset}, E. and {Guy}, L.~P. and {Halbwachs}, J. -L. and
         {Hambly}, N.~C. and {Harrison}, D.~L. and {Hern{\'a}ndez}, J. and
         {Hestroffer}, D. and {Hodgkin}, S.~T. and {Hutton}, A. and
         {Jasniewicz}, G. and {Jean-Antoine-Piccolo}, A. and {Jordan}, S. and
         {Korn}, A.~J. and {Krone-Martins}, A. and {Lanzafame}, A.~C. and
         {Lebzelter}, T. and {L{\"o}ffler}, W. and {Manteiga}, M. and
         {Marrese}, P.~M. and {Mart{\'\i}n-Fleitas}, J.~M. and {Moitinho}, A. and
         {Mora}, A. and {Muinonen}, K. and {Osinde}, J. and {Pancino}, E. and
         {Pauwels}, T. and {Petit}, J. -M. and {Recio-Blanco}, A. and
         {Richards}, P.~J. and {Rimoldini}, L. and {Robin}, A.~C. and
         {Sarro}, L.~M. and {Siopis}, C. and {Smith}, M. and {Sozzetti}, A. and
         {S{\"u}veges}, M. and {Torra}, J. and {van Reeven}, W. and {Abbas}, U. and
         {Abreu Aramburu}, A. and {Accart}, S. and {Aerts}, C. and
         {Altavilla}, G. and {{\'A}lvarez}, M.~A. and {Alvarez}, R. and
         {Alves}, J. and {Anderson}, R.~I. and {Andrei}, A.~H. and
         {Anglada Varela}, E. and {Antiche}, E. and {Antoja}, T. and
         {Arcay}, B. and {Astraatmadja}, T.~L. and {Bach}, N. and
         {Baker}, S.~G. and {Balaguer-N{\'u}{\~n}ez}, L. and {Balm}, P. and
         {Barache}, C. and {Barata}, C. and {Barbato}, D. and {Barblan}, F. and
         {Barklem}, P.~S. and {Barrado}, D. and {Barros}, M. and
         {Barstow}, M.~A. and {Bartholom{\'e} Mu{\~n}oz}, S. and
         {Bassilana}, J. -L. and {Becciani}, U. and {Bellazzini}, M. and
         {Berihuete}, A. and {Bertone}, S. and {Bianchi}, L. and
         {Bienaym{\'e}}, O. and {Blanco-Cuaresma}, S. and {Boch}, T. and
         {Boeche}, C. and {Bombrun}, A. and {Borrachero}, R. and {Bossini}, D. and
         {Bouquillon}, S. and {Bourda}, G. and {Bragaglia}, A. and
         {Bramante}, L. and {Breddels}, M.~A. and {Bressan}, A. and
         {Brouillet}, N. and {Br{\"u}semeister}, T. and {Brugaletta}, E. and
         {Bucciarelli}, B. and {Burlacu}, A. and {Busonero}, D. and
         {Butkevich}, A.~G. and {Buzzi}, R. and {Caffau}, E. and
         {Cancelliere}, R. and {Cannizzaro}, G. and {Cantat-Gaudin}, T. and
         {Carballo}, R. and {Carlucci}, T. and {Carrasco}, J.~M. and
         {Casamiquela}, L. and {Castellani}, M. and {Castro-Ginard}, A. and
         {Charlot}, P. and {Chemin}, L. and {Chiavassa}, A. and {Cocozza}, G. and
         {Costigan}, G. and {Cowell}, S. and {Crifo}, F. and {Crosta}, M. and
         {Crowley}, C. and {Cuypers}, J. and {Dafonte}, C. and {Damerdji}, Y. and
         {Dapergolas}, A. and {David}, P. and {David}, M. and {de Laverny}, P. and
         {De Luise}, F. and {De March}, R. and {de Martino}, D. and
         {de Souza}, R. and {de Torres}, A. and {Debosscher}, J. and
         {del Pozo}, E. and {Delbo}, M. and {Delgado}, A. and {Delgado}, H.~E. and
         {Di Matteo}, P. and {Diakite}, S. and {Diener}, C. and {Distefano}, E. and
         {Dolding}, C. and {Drazinos}, P. and {Dur{\'a}n}, J. and
         {Edvardsson}, B. and {Enke}, H. and {Eriksson}, K. and {Esquej}, P. and
         {Eynard Bontemps}, G. and {Fabre}, C. and {Fabrizio}, M. and
         {Faigler}, S. and {Falc{\~a}o}, A.~J. and {Farr{\`a}s Casas}, M. and
         {Federici}, L. and {Fedorets}, G. and {Fernique}, P. and
         {Figueras}, F. and {Filippi}, F. and {Findeisen}, K. and {Fonti}, A. and
         {Fraile}, E. and {Fraser}, M. and {Fr{\'e}zouls}, B. and {Gai}, M. and
         {Galleti}, S. and {Garabato}, D. and {Garc{\'\i}a-Sedano}, F. and
         {Garofalo}, A. and {Garralda}, N. and {Gavel}, A. and {Gavras}, P. and
         {Gerssen}, J. and {Geyer}, R. and {Giacobbe}, P. and {Gilmore}, G. and
         {Girona}, S. and {Giuffrida}, G. and {Glass}, F. and {Gomes}, M. and
         {Granvik}, M. and {Gueguen}, A. and {Guerrier}, A. and {Guiraud}, J. and
         {Guti{\'e}rrez-S{\'a}nchez}, R. and {Haigron}, R. and
         {Hatzidimitriou}, D. and {Hauser}, M. and {Haywood}, M. and
         {Heiter}, U. and {Helmi}, A. and {Heu}, J. and {Hilger}, T. and
         {Hobbs}, D. and {Hofmann}, W. and {Holland}, G. and {Huckle}, H.~E. and
         {Hypki}, A. and {Icardi}, V. and {Jan{\ss}en}, K. and
         {Jevardat de Fombelle}, G. and {Jonker}, P.~G. and
         {Juh{\'a}sz}, {\'A}. L. and {Julbe}, F. and {Karampelas}, A. and
         {Kewley}, A. and {Klar}, J. and {Kochoska}, A. and {Kohley}, R. and
         {Kolenberg}, K. and {Kontizas}, M. and {Kontizas}, E. and
         {Koposov}, S.~E. and {Kordopatis}, G. and {Kostrzewa-Rutkowska}, Z. and
         {Koubsky}, P. and {Lambert}, S. and {Lanza}, A.~F. and {Lasne}, Y. and
         {Lavigne}, J. -B. and {Le Fustec}, Y. and {Le Poncin-Lafitte}, C. and
         {Lebreton}, Y. and {Leccia}, S. and {Leclerc}, N. and
         {Lecoeur-Taibi}, I. and {Lenhardt}, H. and {Leroux}, F. and {Liao}, S. and
         {Licata}, E. and {Lindstr{\o}m}, H.~E.~P. and {Lister}, T.~A. and
         {Livanou}, E. and {Lobel}, A. and {L{\'o}pez}, M. and {Managau}, S. and
         {Mann}, R.~G. and {Mantelet}, G. and {Marchal}, O. and
         {Marchant}, J.~M. and {Marconi}, M. and {Marinoni}, S. and
         {Marschalk{\'o}}, G. and {Marshall}, D.~J. and {Martino}, M. and
         {Marton}, G. and {Mary}, N. and {Massari}, D. and
         {Matijevi{\v{c}}}, G. and {Mazeh}, T. and {McMillan}, P.~J. and
         {Messina}, S. and {Michalik}, D. and {Millar}, N.~R. and {Molina}, D. and
         {Molinaro}, R. and {Moln{\'a}r}, L. and {Montegriffo}, P. and
         {Mor}, R. and {Morbidelli}, R. and {Morel}, T. and {Morris}, D. and
         {Mulone}, A.~F. and {Muraveva}, T. and {Musella}, I. and
         {Nelemans}, G. and {Nicastro}, L. and {Noval}, L. and {O'Mullane}, W. and
         {Ord{\'e}novic}, C. and {Ord{\'o}{\~n}ez-Blanco}, D. and {Osborne}, P. and
         {Pagani}, C. and {Pagano}, I. and {Pailler}, F. and {Palacin}, H. and
         {Palaversa}, L. and {Panahi}, A. and {Pawlak}, M. and
         {Piersimoni}, A.~M. and {Pineau}, F. -X. and {Plachy}, E. and
         {Plum}, G. and {Poggio}, E. and {Poujoulet}, E. and {Pr{\v{s}}a}, A. and
         {Pulone}, L. and {Racero}, E. and {Ragaini}, S. and {Rambaux}, N. and
         {Ramos-Lerate}, M. and {Regibo}, S. and {Reyl{\'e}}, C. and
         {Riclet}, F. and {Ripepi}, V. and {Riva}, A. and {Rivard}, A. and
         {Rixon}, G. and {Roegiers}, T. and {Roelens}, M. and
         {Romero-G{\'o}mez}, M. and {Rowell}, N. and {Royer}, F. and
         {Ruiz-Dern}, L. and {Sadowski}, G. and {Sagrist{\`a} Sell{\'e}s}, T. and
         {Sahlmann}, J. and {Salgado}, J. and {Salguero}, E. and {Sanna}, N. and
         {Santana-Ros}, T. and {Sarasso}, M. and {Savietto}, H. and
         {Schultheis}, M. and {Sciacca}, E. and {Segol}, M. and
         {Segovia}, J.~C. and {S{\'e}gransan}, D. and {Shih}, I. -C. and
         {Siltala}, L. and {Silva}, A.~F. and {Smart}, R.~L. and {Smith}, K.~W. and
         {Solano}, E. and {Solitro}, F. and {Sordo}, R. and {Soria Nieto}, S. and
         {Souchay}, J. and {Spagna}, A. and {Spoto}, F. and {Stampa}, U. and
         {Steele}, I.~A. and {Steidelm{\"u}ller}, H. and {Stephenson}, C.~A. and
         {Stoev}, H. and {Suess}, F.~F. and {Surdej}, J. and {Szabados}, L. and
         {Szegedi-Elek}, E. and {Tapiador}, D. and {Taris}, F. and {Tauran}, G. and
         {Taylor}, M.~B. and {Teixeira}, R. and {Terrett}, D. and {Teyssand
        ier}, P. and {Thuillot}, W. and {Titarenko}, A. and {Torra Clotet}, F. and
         {Turon}, C. and {Ulla}, A. and {Utrilla}, E. and {Uzzi}, S. and
         {Vaillant}, M. and {Valentini}, G. and {Valette}, V. and
         {van Elteren}, A. and {Van Hemelryck}, E. and {van Leeuwen}, M. and
         {Vaschetto}, M. and {Vecchiato}, A. and {Veljanoski}, J. and
         {Viala}, Y. and {Vicente}, D. and {Vogt}, S. and {von Essen}, C. and
         {Voss}, H. and {Votruba}, V. and {Voutsinas}, S. and {Walmsley}, G. and
         {Weiler}, M. and {Wertz}, O. and {Wevers}, T. and {Wyrzykowski}, {\L}. and
         {Yoldas}, A. and {{\v{Z}}erjal}, M. and {Ziaeepour}, H. and
         {Zorec}, J. and {Zschocke}, S. and {Zucker}, S. and {Zurbach}, C. and
         {Zwitter}, T.},
        title = "{Gaia Data Release 2. Summary of the contents and survey properties}",
      journal = {\aap},
         year = 2018,
        month = aug,
       volume = {616},
          eid = {A1},
        pages = {A1},
          doi = {10.1051/0004-6361/201833051},
archivePrefix = {arXiv},
       eprint = {1804.09365},
 primaryClass = {astro-ph.GA},
       adsurl = {https://ui.adsabs.harvard.edu/abs/2018A&A...616A...1G}
}

@ARTICLE{2024A&A...683A..49Z,
       author = {{Zwintz}, K. and {Pigulski}, A. and {Kuschnig}, R. and {Wade}, G.~A. and {Doherty}, G. and {Earl}, M. and {Lovekin}, C. and {M{\"u}llner}, M. and {Pich{\'e}-Perrier}, S. and {Steindl}, T. and {Beck}, P.~G. and {Bicz}, K. and {Bowman}, D.~M. and {Handler}, G. and {Pablo}, B. and {Popowicz}, A. and {R{\'o}{\.z}a{\'n}ski}, T. and {Miko{\l}ajczyk}, P. and {Baade}, D. and {Koudelka}, O. and {Moffat}, A.~F.~J. and {Neiner}, C. and {Orlea{\'n}ski}, P. and {Smolec}, R. and {Louis}, N. St. and {Weiss}, W.~W. and {Wenger}, M. and {Zoc{\l}o{\'n}ska}, E.},
        title = "{Catalogue of BRITE-Constellation targets. I. Fields 1 to 14 (November 2013-April 2016)}",
      journal = {\aap},
         year = 2024,
        month = mar,
       volume = {683},
          eid = {A49},
        pages = {A49},
          doi = {10.1051/0004-6361/202348236},
archivePrefix = {arXiv},
       eprint = {2311.18382},
 primaryClass = {astro-ph.SR},
       adsurl = {https://ui.adsabs.harvard.edu/abs/2024A&A...683A..49Z}
}

@ARTICLE{2021Univ....7..199W,
       author = {{Weiss}, Werner W. and {Zwintz}, Konstanze and {Kuschnig}, Rainer and {Handler}, Gerald and {Moffat}, Anthony F.~J. and {Baade}, Dietrich and {Bowman}, Dominic M. and {Granzer}, Thomas and {Kallinger}, Thomas and {Koudelka}, Otto F. and {Lovekin}, Catherine C. and {Neiner}, Coralie and {Pablo}, Herbert and {Pigulski}, Andrzej and {Popowicz}, Adam and {Ramiaramanantsoa}, Tahina and {Rucinski}, Slavek M. and {Strassmeier}, Klaus G. and {Wade}, Gregg A.},
        title = "{Space Photometry with BRITE-Constellation}",
      journal = {Universe},
         year = 2021,
        month = jun,
       volume = {7},
       number = {6},
          eid = {199},
        pages = {199},
          doi = {10.3390/universe7060199},
archivePrefix = {arXiv},
       eprint = {2106.12952},
 primaryClass = {astro-ph.SR},
       adsurl = {https://ui.adsabs.harvard.edu/abs/2021Univ....7..199W}
}

@ARTICLE{2007A&A...474..653V,
   author = {{van Leeuwen}, F.},
  journal = {A\&A},
archivePrefix = "arXiv",
     year = 2007,
    month = nov,
   volume = 474,
    pages = {653-664},
      doi = {10.1051/0004-6361:20078357},
   adsurl = {http://adsabs.harvard.edu/abs/2007A%26A...474..653V}
}

@ARTICLE{2023A&A...674A...1G,
       author = {{Gaia Collaboration} and {Vallenari}, A. and {Brown}, A.~G.~A. and {Prusti}, T. and {de Bruijne}, J.~H.~J. and {Arenou}, F. and {Babusiaux}, C. and {Biermann}, M. and {Creevey}, O.~L. and {Ducourant}, C. and {Evans}, D.~W. and {Eyer}, L. and {Guerra}, R. and {Hutton}, A. and {Jordi}, C. and {Klioner}, S.~A. and {Lammers}, U.~L. and {Lindegren}, L. and {Luri}, X. and {Mignard}, F. and {Panem}, C. and {Pourbaix}, D. and {Randich}, S. and {Sartoretti}, P. and {Soubiran}, C. and {Tanga}, P. and {Walton}, N.~A. and {Bailer-Jones}, C.~A.~L. and {Bastian}, U. and {Drimmel}, R. and {Jansen}, F. and {Katz}, D. and {Lattanzi}, M.~G. and {van Leeuwen}, F. and {Bakker}, J. and {Cacciari}, C. and {Casta{\~n}eda}, J. and {De Angeli}, F. and {Fabricius}, C. and {Fouesneau}, M. and {Fr{\'e}mat}, Y. and {Galluccio}, L. and {Guerrier}, A. and {Heiter}, U. and {Masana}, E. and {Messineo}, R. and {Mowlavi}, N. and {Nicolas}, C. and {Nienartowicz}, K. and {Pailler}, F. and {Panuzzo}, P. and {Riclet}, F. and {Roux}, W. and {Seabroke}, G.~M. and {Sordo}, R. and {Th{\'e}venin}, F. and {Gracia-Abril}, G. and {Portell}, J. and {Teyssier}, D. and {Altmann}, M. and {Andrae}, R. and {Audard}, M. and {Bellas-Velidis}, I. and {Benson}, K. and {Berthier}, J. and {Blomme}, R. and {Burgess}, P.~W. and {Busonero}, D. and {Busso}, G. and {C{\'a}novas}, H. and {Carry}, B. and {Cellino}, A. and {Cheek}, N. and {Clementini}, G. and {Damerdji}, Y. and {Davidson}, M. and {de Teodoro}, P. and {Nu{\~n}ez Campos}, M. and {Delchambre}, L. and {Dell'Oro}, A. and {Esquej}, P. and {Fern{\'a}ndez-Hern{\'a}ndez}, J. and {Fraile}, E. and {Garabato}, D. and {Garc{\'\i}a-Lario}, P. and {Gosset}, E. and {Haigron}, R. and {Halbwachs}, J.-L. and {Hambly}, N.~C. and {Harrison}, D.~L. and {Hern{\'a}ndez}, J. and {Hestroffer}, D. and {Hodgkin}, S.~T. and {Holl}, B. and {Jan{\ss}en}, K. and {Jevardat de Fombelle}, G. and {Jordan}, S. and {Krone-Martins}, A. and {Lanzafame}, A.~C. and {L{\"o}ffler}, W. and {Marchal}, O. and {Marrese}, P.~M. and {Moitinho}, A. and {Muinonen}, K. and {Osborne}, P. and {Pancino}, E. and {Pauwels}, T. and {Recio-Blanco}, A. and {Reyl{\'e}}, C. and {Riello}, M. and {Rimoldini}, L. and {Roegiers}, T. and {Rybizki}, J. and {Sarro}, L.~M. and {Siopis}, C. and {Smith}, M. and {Sozzetti}, A. and {Utrilla}, E. and {van Leeuwen}, M. and {Abbas}, U. and {{\'A}brah{\'a}m}, P. and {Abreu Aramburu}, A. and {Aerts}, C. and {Aguado}, J.~J. and {Ajaj}, M. and {Aldea-Montero}, F. and {Altavilla}, G. and {{\'A}lvarez}, M.~A. and {Alves}, J. and {Anders}, F. and {Anderson}, R.~I. and {Anglada Varela}, E. and {Antoja}, T. and {Baines}, D. and {Baker}, S.~G. and {Balaguer-N{\'u}{\~n}ez}, L. and {Balbinot}, E. and {Balog}, Z. and {Barache}, C. and {Barbato}, D. and {Barros}, M. and {Barstow}, M.~A. and {Bartolom{\'e}}, S. and {Bassilana}, J.-L. and {Bauchet}, N. and {Becciani}, U. and {Bellazzini}, M. and {Berihuete}, A. and {Bernet}, M. and {Bertone}, S. and {Bianchi}, L. and {Binnenfeld}, A. and {Blanco-Cuaresma}, S. and {Blazere}, A. and {Boch}, T. and {Bombrun}, A. and {Bossini}, D. and {Bouquillon}, S. and {Bragaglia}, A. and {Bramante}, L. and {Breedt}, E. and {Bressan}, A. and {Brouillet}, N. and {Brugaletta}, E. and {Bucciarelli}, B. and {Burlacu}, A. and {Butkevich}, A.~G. and {Buzzi}, R. and {Caffau}, E. and {Cancelliere}, R. and {Cantat-Gaudin}, T. and {Carballo}, R. and {Carlucci}, T. and {Carnerero}, M.~I. and {Carrasco}, J.~M. and {Casamiquela}, L. and {Castellani}, M. and {Castro-Ginard}, A. and {Chaoul}, L. and {Charlot}, P. and {Chemin}, L. and {Chiaramida}, V. and {Chiavassa}, A. and {Chornay}, N. and {Comoretto}, G. and {Contursi}, G. and {Cooper}, W.~J. and {Cornez}, T. and {Cowell}, S. and {Crifo}, F. and {Cropper}, M. and {Crosta}, M. and {Crowley}, C. and {Dafonte}, C. and {Dapergolas}, A. and {David}, M. and {David}, P. and {de Laverny}, P. and {De Luise}, F. and {De March}, R.},
        title = "{Gaia Data Release 3. Summary of the content and survey properties}",
      journal = {\aap},
         year = 2023,
        month = jun,
       volume = {674},
          eid = {A1},
        pages = {A1},
          doi = {10.1051/0004-6361/202243940},
archivePrefix = {arXiv},
       eprint = {2208.00211},
 primaryClass = {astro-ph.GA},
       adsurl = {https://ui.adsabs.harvard.edu/abs/2023A&A...674A...1G}
}

@ARTICLE{2020A&A...640A..40H,
       author = {{H{\"u}mmerich}, S. and {Paunzen}, E. and {Bernhard}, K.},
        title = "{A plethora of new, magnetic chemically peculiar stars from LAMOST DR4}",
      journal = {\aap},
         year = 2020,
        month = aug,
       volume = {640},
          eid = {A40},
        pages = {A40},
          doi = {10.1051/0004-6361/202037750},
archivePrefix = {arXiv},
       eprint = {2005.14444},
 primaryClass = {astro-ph.SR},
       adsurl = {https://ui.adsabs.harvard.edu/abs/2020A&A...640A..40H}
}

@ARTICLE{2018MNRAS.474.2467H,
       author = {{H{\"u}mmerich}, Stefan and {Niemczura}, Ewa and {Walczak}, Przemys{\l}aw and {Paunzen}, Ernst and {Bernhard}, Klaus and {Murphy}, Simon J. and {Drobek}, Dominik},
        title = "{A spectroscopic and photometric investigation of the mercury-manganese star KIC 6128830}",
      journal = {\mnras},
         year = 2018,
        month = feb,
       volume = {474},
       number = {2},
        pages = {2467-2478},
          doi = {10.1093/mnras/stx2974},
archivePrefix = {arXiv},
       eprint = {1711.08519},
 primaryClass = {astro-ph.SR},
       adsurl = {https://ui.adsabs.harvard.edu/abs/2018MNRAS.474.2467H}
}

@ARTICLE{1968ApJS...17..371L,
       author = {{Lesh}, Janet Rountree},
        title = "{The Kinematics of the Gould Belt: an Expanding Group?}",
      journal = {\apjs},
         year = 1968,
        month = dec,
       volume = {17},
        pages = {371},
          doi = {10.1086/190179},
       adsurl = {https://ui.adsabs.harvard.edu/abs/1968ApJS...17..371L}
}

@ARTICLE{1990ApJ...365..665S,
       author = {{Shore}, Steven N. and {Brown}, Douglas N.},
        title = "{Magnetically Controlled Circumstellar Matter in the Helium-strong Stars}",
      journal = {\apj},
         year = 1990,
        month = dec,
       volume = {365},
        pages = {665},
          doi = {10.1086/169520},
       adsurl = {https://ui.adsabs.harvard.edu/abs/1990ApJ...365..665S}
}

@ARTICLE{2018ApJS..236...16V,
       author = {{VanderPlas}, Jacob T.},
        title = "{Understanding the Lomb-Scargle Periodogram}",
      journal = {\apjs},
         year = 2018,
        month = may,
       volume = {236},
       number = {1},
          eid = {16},
        pages = {16},
          doi = {10.3847/1538-4365/aab766},
archivePrefix = {arXiv},
       eprint = {1703.09824},
 primaryClass = {astro-ph.IM},
       adsurl = {https://ui.adsabs.harvard.edu/abs/2018ApJS..236...16V}
}

@ARTICLE{2006SASS...25...47W,
       author = {{Watson}, C.~L. and {Henden}, A.~A. and {Price}, A.},
        title = "{The International Variable Star Index (VSX)}",
      journal = {Society for Astronomical Sciences Annual Symposium},
         year = 2006,
        month = may,
       volume = {25},
        pages = {47},
       adsurl = {https://ui.adsabs.harvard.edu/abs/2006SASS...25...47W}
}

@book{chatfield2004timeseries,
  address = {Florida, US},
  author = {Chatfield, Chris},
  edition = {6th},
  publisher = {CRC Press},
  title = {The analysis of time series: an introduction},
  year = 2004
}

@INPROCEEDINGS{2015ESS.....350301R,

       author = {{Ricker}, George R.},

        title = "{The Transiting Exoplanet Survey Satellite (TESS): Discovering New Earths and Super-Earths in the Solar Neighborhood}",

    booktitle = {AAS/Division for Extreme Solar Systems Abstracts},

         year = 2015,

       series = {AAS/Division for Extreme Solar Systems Abstracts},

       volume = {47},

        month = dec,

          eid = {503.01},

        pages = {503.01},

       adsurl = {https://ui.adsabs.harvard.edu/abs/2015ESS.....350301R}

}

@ARTICLE{2025AJ....169...48C,
       author = {{Crepp}, Justin R. and {Crass}, Jonathan and {Bechter}, Andrew J. and {Sands}, Brian L. and {Ketterer}, Ryan and {King}, David and {Kopon}, Derek and {Hamper}, Randall and {Engstrom}, Matthew and {Smous}, James E. and {Bechter}, Eric B. and {Harris}, Robert and {Johnson}, Marshall C. and {Baggett}, Nicholas and {Dulz}, Shannon and {Vansickle}, Michael and {Conrad}, Al and {Ertel}, Steve and {Gaudi}, B. Scott and {Hinz}, Philip and {Kuchner}, Marc and {Montoya}, Manny and {Onuma}, Eleanya and {Ott}, Melanie and {Pogge}, Richard and {Rahmer}, Gustavo and {Reynolds}, Robert and {Schwab}, Christian and {Stapelfeldt}, Karl and {Thomes}, Joseph and {Vaz}, Amali and {Wang}, Ji and {Woodward}, Charles E.},
        title = "{Resolving the Young 2 Cygni Runaway Star into a Binary Using iLocater}",
      journal = {\aj},
         year = 2025,
        month = jan,
       volume = {169},
       number = {1},
          eid = {48},
        pages = {48},
          doi = {10.3847/1538-3881/ad9b1d},
archivePrefix = {arXiv},
       eprint = {2412.06982},
 primaryClass = {astro-ph.IM},
       adsurl = {https://ui.adsabs.harvard.edu/abs/2025AJ....169...48C}
}

@BOOK{2009ssc..book.....G,
       author = {{Gray}, Richard O. and {Corbally}, J., Christopher},
        title = "{Stellar Spectral Classification}",
         year = 2009,
       adsurl = {https://ui.adsabs.harvard.edu/abs/2009ssc..book.....G}
}

@ARTICLE{2024A&A...692A.231H,
       author = {{H{\"u}mmerich}, S. and {Bernhard}, K. and {Paunzen}, E.},
        title = "{A new sample of super-slowly rotating Ap (ssrAp) stars from the Zwicky Transient Facility survey}",
      journal = {\aap},
         year = 2024,
        month = dec,
       volume = {692},
          eid = {A231},
        pages = {A231},
          doi = {10.1051/0004-6361/202452075},
archivePrefix = {arXiv},
       eprint = {2411.01534},
 primaryClass = {astro-ph.SR},
       adsurl = {https://ui.adsabs.harvard.edu/abs/2024A&A...692A.231H}
}

@ARTICLE{2020CoSka..50..570P,
       author = {{Paunzen}, E.},
        title = "{Binary fraction of magnetic chemically peculiar stars}",
      journal = {Contributions of the Astronomical Observatory Skalnate Pleso},
         year = 2020,
        month = mar,
       volume = {50},
       number = {2},
        pages = {570-573},
          doi = {10.31577/caosp.2020.50.2.570},
       adsurl = {https://ui.adsabs.harvard.edu/abs/2020CoSka..50..570P}
}

@ARTICLE{2019MNRAS.487.4230S,
       author = {{Skarka}, M. and {Kab{\'a}th}, P. and {Paunzen}, E. and {Fedurco}, M. and {Budaj}, J. and {Dupkala}, D. and {Krti{\v{c}}ka}, J. and {Hatzes}, A. and {Pribulla}, T. and {Parimucha}, {\v{S}}. and {Mikul{\'a}{\v{s}}ek}, Z. and {Guenther}, E. and {Sabotta}, S. and {Bla{\v{z}}ek}, M. and {Dvo{\v{r}}{\'a}kov{\'a}}, J. and {Hamb{\'a}lek}, L. and {Klocov{\'a}}, T. and {Koll{\'a}r}, V. and {Kundra}, E. and {{\v{S}}lechta}, M. and {Va{\~n}ko}, M.},
        title = "{HD 99458: First time ever Ap-type star as a {\ensuremath{\delta}} Scuti pulsator in a short period eclipsing binary?}",
      journal = {\mnras},
         year = 2019,
        month = aug,
       volume = {487},
       number = {3},
        pages = {4230-4237},
          doi = {10.1093/mnras/stz1478},
archivePrefix = {arXiv},
       eprint = {1906.01877},
 primaryClass = {astro-ph.SR},
       adsurl = {https://ui.adsabs.harvard.edu/abs/2019MNRAS.487.4230S}
}

@ARTICLE{2025arXiv250908426S,
       author = {{Southworth}, John and {Bowman}, Dominic},
        title = "{Pulsations in Binary Star Systems}",
      journal = {arXiv e-prints},
         year = 2025,
        month = sep,
          eid = {arXiv:2509.08426},
        pages = {arXiv:2509.08426},
          doi = {10.48550/arXiv.2509.08426},
archivePrefix = {arXiv},
       eprint = {2509.08426},
 primaryClass = {astro-ph.SR},
       adsurl = {https://ui.adsabs.harvard.edu/abs/2025arXiv250908426S}
}

@ARTICLE{2022MNRAS.514.3485G,
       author = {{Giarrusso}, M. and {Cecconi}, M. and {Cosentino}, R. and {Munari}, M. and {Ghedina}, A. and {Ambrosino}, F. and {Boschin}, W. and {Leone}, F.},
        title = "{Twenty-year monitoring of the surface magnetic fields of chemically peculiar stars}",
      journal = {\mnras},
         year = 2022,
        month = aug,
       volume = {514},
       number = {3},
        pages = {3485-3509},
          doi = {10.1093/mnras/stac1488},
archivePrefix = {arXiv},
       eprint = {2108.12527},
 primaryClass = {astro-ph.SR},
       adsurl = {https://ui.adsabs.harvard.edu/abs/2022MNRAS.514.3485G}
}

@ARTICLE{2025A&A...703A.102M,
       author = {{Mathys}, G. and {Holdsworth}, D.~L. and {Giarrusso}, M. and {Kurtz}, D.~W. and {Catanzaro}, G. and {Leone}, F.},
        title = "{Super-slowly rotating Ap (ssrAp) stars: New spectroscopic observations}",
      journal = {\aap},
         year = 2025,
        month = nov,
       volume = {703},
          eid = {A102},
        pages = {A102},
          doi = {10.1051/0004-6361/202556104},
archivePrefix = {arXiv},
       eprint = {2509.05457},
 primaryClass = {astro-ph.SR},
       adsurl = {https://ui.adsabs.harvard.edu/abs/2025A&A...703A.102M}
}

@ARTICLE{1987ApJ...323..325B,
       author = {{Bohlender}, David A. and {Brown}, Douglas N. and {Landstreet}, J.~D. and {Thompson}, Ian B.},
        title = "{Magnetic Field Measurements of Helium-strong Stars}",
      journal = {\apj},
         year = 1987,
        month = dec,
       volume = {323},
        pages = {325},
          doi = {10.1086/165830},
       adsurl = {https://ui.adsabs.harvard.edu/abs/1987ApJ...323..325B}
}

@ARTICLE{1983ApJS...53..151B,
       author = {{Borra}, E.~F. and {Landstreet}, J.~D. and {Thompson}, I.},
        title = "{The magnetic fields of the helium-weak B stars.}",
      journal = {\apjs},
         year = 1983,
        month = sep,
       volume = {53},
        pages = {151-167},
          doi = {10.1086/190889},
       adsurl = {https://ui.adsabs.harvard.edu/abs/1983ApJS...53..151B}
}

@ARTICLE{2023ApJS..268....4F,
       author = {{Fetherolf}, Tara and {Pepper}, Joshua and {Simpson}, Emilie and {Kane}, Stephen R. and {Mo{\v{c}}nik}, Teo and {English}, John Edward and {Antoci}, Victoria and {Huber}, Daniel and {Jenkins}, Jon M. and {Stassun}, Keivan and {Twicken}, Joseph D. and {Vanderspek}, Roland and {Winn}, Joshua N.},
        title = "{Variability Catalog of Stars Observed during the TESS Prime Mission}",
      journal = {\apjs},
         year = 2023,
        month = sep,
       volume = {268},
       number = {1},
          eid = {4},
        pages = {4},
          doi = {10.3847/1538-4365/acdee5},
archivePrefix = {arXiv},
       eprint = {2208.11721},
 primaryClass = {astro-ph.SR},
       adsurl = {https://ui.adsabs.harvard.edu/abs/2023ApJS..268....4F}
}

@ARTICLE{2024A&A...688A..62K,
       author = {{Kallinger}, T. and {Weiss}, W.~W. and {Kuschnig}, R. and {Stassun}, K.~G.},
        title = "{A benchmark rapidly oscillating chemically peculiar (roAp) star: {\ensuremath{\alpha}} Cir}",
      journal = {\aap},
         year = 2024,
        month = aug,
       volume = {688},
          eid = {A62},
        pages = {A62},
          doi = {10.1051/0004-6361/202346942},
       adsurl = {https://ui.adsabs.harvard.edu/abs/2024A&A...688A..62K}
}

@ARTICLE{2019MNRAS.483.3127S,
       author = {{Sikora}, J. and {Wade}, G.~A. and {Power}, J. and {Neiner}, C.},
        title = "{A volume-limited survey of mCP stars within 100 pc II: rotational and magnetic properties}",
      journal = {\mnras},
         year = 2019,
        month = mar,
       volume = {483},
       number = {3},
        pages = {3127-3145},
          doi = {10.1093/mnras/sty2895},
archivePrefix = {arXiv},
       eprint = {1811.05635},
 primaryClass = {astro-ph.SR},
       adsurl = {https://ui.adsabs.harvard.edu/abs/2019MNRAS.483.3127S}
}

@ARTICLE{2021MNRAS.506.5328K,
       author = {{Kochukhov}, O. and {Khalack}, V. and {Kobzar}, O. and {Neiner}, C. and {Paunzen}, E. and {Labadie-Bartz}, J. and {David-Uraz}, A.},
        title = "{TESS survey of rotational and pulsational variability of mercury-manganese stars}",
      journal = {\mnras},
         year = 2021,
        month = oct,
       volume = {506},
       number = {4},
        pages = {5328-5344},
          doi = {10.1093/mnras/stab2107},
archivePrefix = {arXiv},
       eprint = {2107.09096},
 primaryClass = {astro-ph.SR},
       adsurl = {https://ui.adsabs.harvard.edu/abs/2021MNRAS.506.5328K}
}




\appendix

\section{The used data sets from the BRITE mission}

Here we list the individual data sets used for this work.

\begin{table*}
    \centering
    \caption{The used individual data sets with the field identification (Field), setup number (Mode), and BRITE satellite (Sat).}
    \label{obs_log}
    \begin{adjustbox}{max width=\textwidth}
    \begin{tabular}{cccccccccccccc}
    \hline
    HD & Field & HJD(start) & HJD(end) & $N_\mathrm{Obs}$ & Mode & Sat & HD & Field & HJD(start) & HJD(end) & $N_\mathrm{Obs}$ & Mode & Sat \\
        \hline
6961	&	11	&	7262.427009	&	7321.573291	&	19619	&	setup1	&	BAb	&	28319	&	40	&	8430.592994	&	8435.986381	&	1099	&	setup1	&	BTr	\\
6961	&	19	&	7608.265374	&	7631.152040	&	2453	&	setup1	&	BAb	&	28319	&	40	&	8436.051669	&	8557.971099	&	36369	&	setup2	&	BTr	\\
6961	&	19	&	7624.383151	&	7754.152064	&	23243	&	setup2	&	BAb	&	28319	&	48	&	8749.881462	&	8802.654118	&	24378	&	setup1	&	BLb	\\
6961	&	19	&	7645.067485	&	7645.983539	&	571	&	setup1	&	UBr	&	28319	&	48	&	8802.654823	&	8871.040905	&	29800	&	setup2	&	BLb	\\
6961	&	19	&	7646.392592	&	7786.139672	&	30747	&	setup2	&	UBr	&	28319	&	48	&	8876.494361	&	8909.346688	&	8811	&	setup3	&	BLb	\\
6961	&	30	&	7973.350910	&	8133.594702	&	12932	&	setup2	&	BAb	&	28319	&	48	&	8749.598760	&	8823.564787	&	24981	&	setup1	&	BTr	\\
6961	&	39	&	8339.259892	&	8436.818206	&	4283	&	setup1	&	BAb	&	28319	&	48	&	8823.565498	&	8891.460338	&	24225	&	setup2	&	BTr	\\
11415	&	11	&	7263.407064	&	7321.572737	&	18091	&	setup1	&	BAb	&	28319	&	58	&	9129.535738	&	9163.453562	&	13857	&	setup1	&	BTr	\\
11415	&	19	&	7608.264463	&	7631.151532	&	2291	&	setup1	&	BAb	&	28319	&	58	&	9163.514825	&	9224.655019	&	21967	&	setup2	&	BTr	\\
11415	&	19	&	7624.453561	&	7753.456064	&	23459	&	setup2	&	BAb	&	28319	&	58	&	9224.655731	&	9289.395670	&	24418	&	setup3	&	BTr	\\
11415	&	19	&	7645.066538	&	7645.982595	&	565	&	setup1	&	UBr	&	28527	&	31	&	8030.083751	&	8057.480187	&	9710	&	setup1	&	BLb	\\
11415	&	19	&	7646.391651	&	7786.140540	&	30215	&	setup2	&	UBr	&	28527	&	31	&	8057.609912	&	8098.075660	&	18428	&	setup2	&	BLb	\\
11415	&	30	&	7973.350000	&	8133.595403	&	12760	&	setup2	&	BAb	&	28527	&	31	&	8032.708853	&	8033.599698	&	195	&	setup3	&	BTr	\\
11415	&	39	&	8339.258977	&	8436.817712	&	4176	&	setup1	&	BAb	&	28527	&	31	&	8033.732862	&	8053.531351	&	2723	&	setup4	&	BTr	\\
15633	&	21	&	7701.451071	&	7720.403617	&	7602	&	setup2	&	BHr	&	28527	&	31	&	8053.801238	&	8109.485643	&	6382	&	setup5	&	BTr	\\
15633	&	21	&	7720.473307	&	7756.876112	&	16321	&	setup3	&	BHr	&	28527	&	40	&	8375.228572	&	8552.146662	&	7429	&	setup1	&	BAb	\\
18296	&	5	&	6917.226638	&	6926.159313	&	1600	&	setup3	&	BAb	&	29140	&	31	&	8057.609816	&	8098.075575	&	18474	&	setup2	&	BLb	\\
18296	&	5	&	6904.510144	&	6922.587607	&	10232	&	setup2	&	UBr	&	29140	&	31	&	8098.135499	&	8129.471117	&	14683	&	setup3	&	BLb	\\
18866	&	56	&	9081.428651	&	9100.852266	&	5321	&	setup3	&	BHr	&	29140	&	31	&	8129.537983	&	8171.354441	&	15392	&	setup4	&	BLb	\\
18866	&	56	&	9100.978758	&	9107.320971	&	2116	&	setup4	&	BHr	&	29140	&	31	&	8165.696382	&	8179.615694	&	6133	&	setup10	&	BTr	\\
20320	&	21	&	7668.067231	&	7701.385170	&	8139	&	setup1	&	BHr	&	29140	&	31	&	8109.611147	&	8130.023953	&	9788	&	setup6	&	BTr	\\
22634	&	56	&	9081.428270	&	9100.851864	&	5787	&	setup3	&	BHr	&	29140	&	31	&	8133.834332	&	8165.570581	&	18687	&	setup7	&	BTr	\\
22920	&	21	&	7668.066905	&	7701.384764	&	8146	&	setup1	&	BHr	&	29140	&	48	&	8749.881351	&	8802.654009	&	24371	&	setup1	&	BLb	\\
22920	&	21	&	7701.450829	&	7720.404825	&	7678	&	setup2	&	BHr	&	29140	&	48	&	8802.654714	&	8871.040917	&	29796	&	setup2	&	BLb	\\
22920	&	21	&	7720.473576	&	7756.877701	&	16346	&	setup3	&	BHr	&	29305	&	56	&	9052.913719	&	9053.326363	&	208	&	setup1	&	BHr	\\
23408	&	5	&	6913.529004	&	6917.160531	&	618	&	setup1	&	BAb	&	29305	&	56	&	9053.384318	&	9081.301609	&	11065	&	setup2	&	BHr	\\
23408	&	5	&	6994.597283	&	7045.151783	&	1802	&	setup2	&	BAb	&	29305	&	56	&	9081.427398	&	9100.851358	&	4943	&	setup3	&	BHr	\\
23408	&	5	&	6917.225970	&	6926.158737	&	1623	&	setup3	&	BAb	&	29305	&	56	&	9100.977148	&	9107.319967	&	2164	&	setup4	&	BHr	\\
23408	&	5	&	6937.464514	&	6959.514965	&	11970	&	setup4	&	BAb	&	29388	&	31	&	8057.609802	&	8098.075631	&	18468	&	setup2	&	BLb	\\
23408	&	5	&	6960.404487	&	6994.113214	&	8978	&	setup5	&	BAb	&	30780	&	31	&	8032.708410	&	8033.599260	&	191	&	setup3	&	BTr	\\
23408	&	5	&	6904.509372	&	6922.586992	&	10274	&	setup2	&	UBr	&	30780	&	31	&	8033.732424	&	8053.531042	&	2663	&	setup4	&	BTr	\\
23408	&	5	&	6925.019415	&	6993.457482	&	62686	&	setup3	&	UBr	&	30780	&	31	&	8053.800931	&	8109.485828	&	6413	&	setup5	&	BTr	\\
23408	&	5	&	6996.588093	&	7072.408486	&	45034	&	setup6	&	UBr	&	32549	&	31	&	8030.083027	&	8057.479710	&	9720	&	setup1	&	BLb	\\
23408	&	5	&	6996.592170	&	7072.408131	&	28495	&	setup7	&	UBr	&	32549	&	31	&	8098.135647	&	8129.471636	&	14681	&	setup3	&	BLb	\\
23850	&	5	&	6913.528941	&	6917.160470	&	626	&	setup1	&	BAb	&	32549	&	31	&	8129.538503	&	8171.355202	&	16337	&	setup4	&	BLb	\\
23850	&	5	&	6994.597298	&	7045.153245	&	2717	&	setup2	&	BAb	&	32549	&	48	&	8749.880636	&	8802.653810	&	24321	&	setup1	&	BLb	\\
23850	&	5	&	6917.225909	&	6926.158681	&	1639	&	setup3	&	BAb	&	32549	&	48	&	8802.654515	&	8871.041534	&	29784	&	setup2	&	BLb	\\
23850	&	5	&	6937.464467	&	6959.515113	&	11397	&	setup4	&	BAb	&	33641	&	20	&	7672.147483	&	7697.065640	&	10937	&	setup4	&	BTr	\\
23850	&	5	&	6960.404463	&	6994.113229	&	9038	&	setup5	&	BAb	&	33641	&	20	&	7697.191926	&	7730.909267	&	9610	&	setup5	&	BTr	\\
23850	&	5	&	6904.509306	&	6922.586934	&	10283	&	setup2	&	UBr	&	33904	&	6	&	6937.368753	&	6944.987580	&	1672	&	setup3	&	BAb	\\
23850	&	5	&	6925.019358	&	6993.457497	&	64865	&	setup3	&	UBr	&	33904	&	6	&	6945.390015	&	6972.046255	&	2447	&	setup4	&	BAb	\\
23850	&	5	&	6996.588110	&	7072.408553	&	62564	&	setup6	&	UBr	&	33904	&	6	&	6924.717259	&	6972.123821	&	26932	&	setup1	&	BTr	\\
23850	&	5	&	6996.592542	&	7072.408376	&	27853	&	setup7	&	UBr	&	33904	&	6	&	6972.168888	&	6984.746523	&	1270	&	setup2	&	BTr	\\
25823	&	5	&	6904.508843	&	6922.586509	&	10214	&	setup2	&	UBr	&	33959	&	20	&	7645.831857	&	7646.957252	&	791	&	setup1	&	BLb	\\
26961	&	20	&	7672.147922	&	7697.065687	&	10864	&	setup4	&	BTr	&	33959	&	20	&	7655.100808	&	7792.588287	&	23181	&	setup3	&	BLb	\\
26961	&	20	&	7697.191970	&	7730.980673	&	9575	&	setup5	&	BTr	&	33959	&	20	&	7792.648208	&	7819.768046	&	13951	&	setup4	&	BLb	\\
26961	&	20	&	7731.180780	&	7822.683538	&	26809	&	setup7	&	BTr	&	33959	&	20	&	7646.555013	&	7672.086048	&	7889	&	setup2	&	BTr	\\
27463	&	56	&	9053.384883	&	9081.301971	&	11075	&	setup2	&	BHr	&	33959	&	20	&	7672.147546	&	7697.065743	&	10928	&	setup4	&	BTr	\\
27463	&	56	&	9081.427759	&	9100.851527	&	5875	&	setup3	&	BHr	&	33959	&	20	&	7697.192029	&	7730.846705	&	9606	&	setup5	&	BTr	\\
27463	&	56	&	9100.977315	&	9107.320067	&	2165	&	setup4	&	BHr	&	33959	&	20	&	7731.181394	&	7822.684311	&	26828	&	setup7	&	BTr	\\
27628	&	31	&	8109.611111	&	8130.023823	&	9693	&	setup6	&	BTr	&	34452	&	20	&	7672.147464	&	7697.065678	&	10949	&	setup4	&	BTr	\\
27962	&	31	&	8030.083838	&	8057.480247	&	9697	&	setup1	&	BLb	&	35039	&	6	&	6987.605537	&	6995.591762	&	3855	&	setup3	&	BTr	\\
28319	&	31	&	8030.083792	&	8057.480210	&	9728	&	setup1	&	BLb	&	35039	&	40	&	8399.986813	&	8468.663122	&	29512	&	setup1	&	BLb	\\
28319	&	31	&	8057.609936	&	8098.075649	&	18437	&	setup2	&	BLb	&	35039	&	40	&	8468.663615	&	8540.371234	&	32811	&	setup2	&	BLb	\\
28319	&	31	&	8098.135573	&	8129.471128	&	14677	&	setup3	&	BLb	&	35039	&	49	&	8745.169494	&	8917.628042	&	21688	&	setup1	&	BAb	\\
28319	&	31	&	8129.537995	&	8171.354368	&	15273	&	setup4	&	BLb	&	35039	&	59	&	9111.337886	&	9256.118897	&	9534	&	setup1	&	BAb	\\
28319	&	31	&	8165.696319	&	8179.615607	&	5926	&	setup10	&	BTr	&	35039	&	65	&	9469.268275	&	9641.156269	&	11205	&	setup1	&	BAb	\\
28319	&	31	&	8032.709129	&	8033.599738	&	185	&	setup3	&	BTr	&	35497	&	31	&	8030.082469	&	8057.479324	&	9321	&	setup1	&	BLb	\\
28319	&	31	&	8033.732901	&	8053.531378	&	2601	&	setup4	&	BTr	&	35497	&	31	&	8057.609052	&	8098.075733	&	18457	&	setup2	&	BLb	\\
28319	&	31	&	8053.801264	&	8109.485623	&	6152	&	setup5	&	BTr	&	35497	&	31	&	8098.135659	&	8129.471960	&	14668	&	setup3	&	BLb	\\
28319	&	31	&	8109.611200	&	8130.023964	&	9442	&	setup6	&	BTr	&	35497	&	31	&	8129.538828	&	8171.354605	&	16250	&	setup4	&	BLb	\\
28319	&	31	&	8133.834334	&	8165.570045	&	17932	&	setup7	&	BTr	&	35497	&	31	&	8032.707600	&	8033.598456	&	191	&	setup3	&	BTr	\\
28319	&	40	&	8375.228856	&	8552.146614	&	18395	&	setup1	&	BAb	&	35497	&	31	&	8130.150871	&	8131.461390	&	953	&	setup8	&	BTr	\\
        \hline
    \end{tabular}
    \end{adjustbox}   
\end{table*}

\begin{table*}
    \centering
    \caption{Table \ref{obs_log} continued.}
    \begin{adjustbox}{max width=\textwidth}
    \begin{tabular}{cccccccccccccc}
    \hline
    HD & Field & HJD(start) & HJD(end) & $N_\mathrm{Obs}$ & Mode & Sat & HD & Field & HJD(start) & HJD(end) & $N_\mathrm{Obs}$ & Mode & Sat \\
        \hline
    35497	&	31	&	8131.651774	&	8179.425844	&	9836	&	setup9	&	BTr	&	67523	&	12	&	7322.154950	&	7351.083088	&	9892	&	setup3	&	BLb	\\
35497	&	48	&	8749.880043	&	8802.653555	&	24358	&	setup1	&	BLb	&	67523	&	12	&	7351.212355	&	7359.175012	&	3532	&	setup4	&	BLb	\\
35497	&	48	&	8802.654260	&	8871.041950	&	29779	&	setup2	&	BLb	&	67523	&	12	&	7359.239769	&	7419.154317	&	18507	&	setup5	&	BLb	\\
36960	&	6	&	6998.561177	&	7049.170103	&	6310	&	setup5	&	BHr	&	67523	&	12	&	7419.160197	&	7488.670047	&	13824	&	setup5	&	BLb	\\
36960	&	6	&	7049.566796	&	7056.862852	&	4035	&	setup6	&	BHr	&	67523	&	12	&	7367.186888	&	7497.257714	&	43410	&	setup3	&	BTr	\\
36960	&	6	&	7056.987847	&	7095.514025	&	21855	&	setup7	&	BHr	&	67523	&	41	&	8426.372368	&	8435.282254	&	3011	&	setup1	&	BHr	\\
36960	&	6	&	6998.522982	&	7043.565234	&	5141	&	setup3	&	BLb	&	67523	&	41	&	8435.408103	&	8491.780591	&	20543	&	setup2	&	BHr	\\
36960	&	6	&	7052.754099	&	7098.281604	&	27834	&	setup6	&	BLb	&	67523	&	41	&	8491.838082	&	8564.379496	&	15383	&	setup3	&	BHr	\\
40183	&	20	&	7655.099871	&	7792.588827	&	22557	&	setup3	&	BLb	&	73634	&	7	&	7003.331560	&	7005.288793	&	310	&	setup1	&	BAb	\\
40183	&	20	&	7792.648749	&	7819.768890	&	13931	&	setup4	&	BLb	&	73634	&	7	&	7005.353693	&	7011.221903	&	1267	&	setup2	&	BAb	\\
40183	&	20	&	7646.547082	&	7647.095395	&	122	&	setup1	&	BTr	&	73634	&	7	&	7011.356370	&	7036.181932	&	4303	&	setup3	&	BAb	\\
40183	&	20	&	7646.554117	&	7672.083905	&	5621	&	setup2	&	BTr	&	73634	&	7	&	7010.757728	&	7028.112183	&	4061	&	setup1	&	BTr	\\
40183	&	20	&	7647.229993	&	7730.854828	&	22116	&	setup3	&	BTr	&	73634	&	7	&	7028.166723	&	7035.551748	&	2428	&	setup2	&	BTr	\\
40183	&	20	&	7731.192831	&	7822.693743	&	14779	&	setup6	&	BTr	&	73634	&	7	&	7035.612892	&	7080.135274	&	11031	&	setup3	&	BTr	\\
40292	&	23	&	7697.216793	&	7702.066788	&	1655	&	setup1	&	BTr	&	73634	&	25	&	7808.284665	&	7833.252121	&	9136	&	setup2	&	BAb	\\
40292	&	23	&	7702.128329	&	7721.586201	&	7182	&	setup2	&	BTr	&	73634	&	66	&	9552.415480	&	9559.377638	&	2053	&	setup1	&	BTr	\\
40292	&	23	&	7721.649389	&	7731.141477	&	3604	&	setup3	&	BTr	&	73634	&	66	&	9559.440878	&	9723.348773	&	62694	&	setup2	&	BTr	\\
40312	&	20	&	7645.830928	&	7646.956321	&	791	&	setup1	&	BLb	&	74560	&	7	&	7005.352926	&	7011.221084	&	1285	&	setup2	&	BAb	\\
40312	&	20	&	7655.099871	&	7792.588981	&	23317	&	setup3	&	BLb	&	81188	&	7	&	7003.330427	&	7005.287469	&	301	&	setup1	&	BAb	\\
40312	&	20	&	7792.648903	&	7819.768947	&	13962	&	setup4	&	BLb	&	81188	&	7	&	7005.352368	&	7011.220553	&	1267	&	setup2	&	BAb	\\
40312	&	20	&	7646.547518	&	7647.095363	&	113	&	setup1	&	BTr	&	81188	&	7	&	7011.355020	&	7036.180635	&	4354	&	setup3	&	BAb	\\
40312	&	20	&	7646.554083	&	7672.085156	&	7756	&	setup2	&	BTr	&	81188	&	7	&	7036.248394	&	7058.769850	&	6551	&	setup4	&	BAb	\\
40312	&	20	&	7672.146655	&	7697.065054	&	10911	&	setup4	&	BTr	&	81188	&	7	&	7059.464128	&	7074.035006	&	4928	&	setup5	&	BAb	\\
40312	&	20	&	7697.191341	&	7730.846483	&	9526	&	setup5	&	BTr	&	81188	&	7	&	7072.991029	&	7091.194156	&	4978	&	setup6	&	BAb	\\
40312	&	20	&	7731.181177	&	7822.685222	&	26286	&	setup7	&	BTr	&	81188	&	7	&	7091.538915	&	7166.153780	&	18461	&	setup7	&	BAb	\\
53244	&	12	&	7340.761523	&	7344.122485	&	2702	&	setup1	&	BTr	&	81188	&	7	&	7010.755872	&	7028.110840	&	4222	&	setup1	&	BTr	\\
53244	&	12	&	7344.173667	&	7367.126456	&	17992	&	setup2	&	BTr	&	81188	&	7	&	7028.165381	&	7035.550446	&	2410	&	setup2	&	BTr	\\
54118	&	61	&	9289.537997	&	9318.400940	&	10409	&	setup1	&	BTr	&	81188	&	7	&	7035.611591	&	7080.134636	&	10863	&	setup3	&	BTr	\\
54118	&	61	&	9318.466931	&	9345.487121	&	8447	&	setup2	&	BTr	&	81188	&	7	&	7080.192061	&	7091.804925	&	6942	&	setup4	&	BTr	\\
55719	&	33	&	8103.396381	&	8160.918002	&	30078	&	setup4	&	BHr	&	81188	&	7	&	7092.203898	&	7170.036167	&	49885	&	setup5	&	BTr	\\
56022	&	8	&	7097.990326	&	7108.262977	&	4207	&	setup3	&	BHr	&	81188	&	14	&	7374.306217	&	7418.792623	&	2213	&	setup1	&	BAb	\\
56022	&	23	&	7760.579041	&	7784.436136	&	15966	&	setup1	&	BHr	&	81188	&	14	&	7375.284689	&	7384.285186	&	1510	&	setup2	&	BAb	\\
56022	&	23	&	7739.230726	&	7746.084681	&	3771	&	setup2	&	BLb	&	81188	&	25	&	7798.661983	&	7808.218499	&	1978	&	setup1	&	BAb	\\
56022	&	23	&	7746.148247	&	7792.562554	&	22209	&	setup3	&	BLb	&	81188	&	25	&	7798.587644	&	7808.220145	&	3629	&	setup1	&	BAb	\\
56022	&	23	&	7792.622478	&	7845.057404	&	27876	&	setup4	&	BLb	&	81188	&	25	&	7808.284449	&	7833.251997	&	6201	&	setup2	&	BAb	\\
56022	&	23	&	7697.216154	&	7702.066026	&	1005	&	setup1	&	BTr	&	81188	&	25	&	7808.283978	&	7833.251997	&	9136	&	setup2	&	BAb	\\
56022	&	23	&	7702.127804	&	7721.587099	&	7078	&	setup2	&	BTr	&	81188	&	25	&	7833.524772	&	7871.107156	&	6562	&	setup3	&	BAb	\\
56022	&	61	&	9162.922098	&	9302.381956	&	15804	&	setup2	&	BHr	&	81188	&	25	&	7833.524302	&	7871.107391	&	11534	&	setup3	&	BAb	\\
56455	&	8	&	7097.990301	&	7108.262980	&	4279	&	setup3	&	BHr	&	81188	&	34	&	8136.551248	&	8247.964652	&	13926	&	setup1	&	BAb	\\
56455	&	8	&	7108.594841	&	7115.750439	&	3725	&	setup4	&	BHr	&	81188	&	34	&	8094.384052	&	8162.231903	&	4996	&	setup1	&	UBr	\\
59635	&	23	&	7697.217094	&	7702.066313	&	1611	&	setup1	&	BTr	&	81188	&	34	&	8162.506298	&	8267.162615	&	21630	&	setup2	&	UBr	\\
59635	&	23	&	7702.127856	&	7721.586416	&	7184	&	setup2	&	BTr	&	81188	&	51	&	8817.666393	&	8853.015019	&	2812	&	setup1	&	BAb	\\
59635	&	61	&	9289.538940	&	9318.401086	&	10628	&	setup1	&	BTr	&	81188	&	53	&	8878.591727	&	8958.519985	&	25139	&	setup1	&	BLb	\\
59635	&	61	&	9318.467076	&	9345.215206	&	8363	&	setup2	&	BTr	&	81188	&	60	&	9275.641622	&	9319.149621	&	1081	&	setup1	&	BAb	\\
64740	&	7	&	7005.353674	&	7011.221790	&	1283	&	setup2	&	BAb	&	82434	&	7	&	7003.330775	&	7005.288391	&	315	&	setup1	&	BAb	\\
64740	&	7	&	7011.356255	&	7036.181444	&	4383	&	setup3	&	BAb	&	82434	&	7	&	7005.353113	&	7011.221401	&	1249	&	setup2	&	BAb	\\
64740	&	7	&	7036.249201	&	7058.770134	&	6674	&	setup4	&	BAb	&	82434	&	7	&	7011.355871	&	7036.181813	&	4275	&	setup3	&	BAb	\\
64740	&	7	&	7059.464395	&	7074.034906	&	4982	&	setup5	&	BAb	&	82434	&	7	&	7010.756713	&	7028.111934	&	4165	&	setup1	&	BTr	\\
64740	&	7	&	7072.990424	&	7091.193812	&	6371	&	setup6	&	BAb	&	82434	&	7	&	7028.166475	&	7035.551619	&	2395	&	setup2	&	BTr	\\
64740	&	7	&	7091.538384	&	7166.152312	&	22110	&	setup7	&	BAb	&	82434	&	7	&	7035.612764	&	7080.135836	&	11214	&	setup3	&	BTr	\\
64740	&	7	&	7010.757115	&	7028.111809	&	4048	&	setup1	&	BTr	&	93030	&	15	&	7423.402068	&	7443.062418	&	697	&	setup1	&	BAb	\\
64740	&	7	&	7028.166348	&	7035.551268	&	2441	&	setup2	&	BTr	&	93030	&	15	&	7443.338025	&	7448.227709	&	824	&	setup2	&	BAb	\\
64740	&	7	&	7035.612411	&	7080.134383	&	11049	&	setup3	&	BTr	&	93030	&	15	&	7448.288262	&	7470.188063	&	4587	&	setup3	&	BAb	\\
64740	&	7	&	7092.203352	&	7170.028682	&	50990	&	setup5	&	BTr	&	93030	&	15	&	7438.046548	&	7506.498153	&	11370	&	setup1	&	BHr	\\
64740	&	8	&	7097.990764	&	7108.263529	&	3854	&	setup3	&	BHr	&	93030	&	15	&	7438.168775	&	7450.477178	&	1998	&	setup1	&	BLb	\\
64740	&	8	&	7108.595393	&	7115.751219	&	3694	&	setup4	&	BHr	&	93030	&	15	&	7457.250932	&	7526.082415	&	680	&	setup2	&	BLb	\\
64740	&	14	&	7374.307481	&	7418.793977	&	2207	&	setup1	&	BAb	&	93030	&	15	&	7526.493228	&	7549.251196	&	3673	&	setup3	&	BLb	\\
64740	&	14	&	7375.285941	&	7384.286312	&	1504	&	setup2	&	BAb	&	93030	&	15	&	7550.212398	&	7553.126325	&	501	&	setup4	&	BLb	\\
64740	&	25	&	7798.587695	&	7808.219952	&	3629	&	setup1	&	BAb	&	93030	&	15	&	7553.187078	&	7555.130631	&	396	&	setup5	&	BLb	\\
64740	&	66	&	9559.440666	&	9723.348484	&	62782	&	setup2	&	BTr	&	93030	&	15	&	7555.192795	&	7585.418404	&	1971	&	setup6	&	BLb	\\
67523	&	12	&	7323.346468	&	7350.269143	&	8850	&	setup3	&	BHr	&	93030	&	15	&	7430.048050	&	7441.993531	&	631	&	setup2	&	BTr	\\
67523	&	12	&	7350.327572	&	7414.884146	&	19731	&	setup4	&	BHr	&	93030	&	15	&	7500.606393	&	7557.167207	&	12714	&	setup4	&	UBr	\\
67523	&	12	&	7408.927467	&	7490.673447	&	23869	&	setup4	&	BHr	&	93030	&	15	&	7558.557296	&	7592.457344	&	6706	&	setup5	&	UBr	\\
        \hline
    \end{tabular}
    \end{adjustbox} 
\end{table*}

\begin{table*}
    \centering
    \caption{Table \ref{obs_log} continued.}
    \begin{adjustbox}{max width=\textwidth}
    \begin{tabular}{cccccccccccccc}
    \hline
    HD & Field & HJD(start) & HJD(end) & $N_\mathrm{Obs}$ & Mode & Sat & HD & Field & HJD(start) & HJD(end) & $N_\mathrm{Obs}$ & Mode & Sat \\
        \hline
    93030	&	43	&	8512.666532	&	8611.327093	&	14162	&	setup1	&	BTr	&	141556	&	9	&	7102.080628	&	7154.669547	&	16514	&	setup1	&	UBr	\\
93030	&	43	&	8611.327801	&	8681.544297	&	17510	&	setup2	&	BTr	&	141556	&	9	&	7155.077952	&	7184.990433	&	13253	&	setup2	&	UBr	\\
104671	&	15	&	7453.888862	&	7454.453089	&	263	&	setup1	&	UBr	&	141556	&	9	&	7185.050505	&	7224.113800	&	14730	&	setup3	&	UBr	\\
104671	&	15	&	7454.516332	&	7472.443574	&	7238	&	setup2	&	UBr	&	141556	&	9	&	7225.012801	&	7264.139671	&	14011	&	setup4	&	UBr	\\
104671	&	15	&	7474.389201	&	7500.404116	&	6451	&	setup3	&	UBr	&	141891	&	18	&	7611.178496	&	7624.140130	&	343	&	setup3	&	BTr	\\
104671	&	62	&	9306.958344	&	9314.448010	&	3869	&	setup3	&	BHr	&	142990	&	44	&	8568.411186	&	8610.351800	&	12346	&	setup1	&	BHr	\\
104671	&	62	&	9314.583660	&	9443.514818	&	16288	&	setup4	&	BHr	&	142990	&	44	&	8610.410227	&	8661.650491	&	23838	&	setup2	&	BHr	\\
104671	&	62	&	9305.538625	&	9449.670247	&	16891	&	setup3	&	BTr	&	142990	&	44	&	8661.910751	&	8723.326656	&	26748	&	setup3	&	BHr	\\
109026	&	15	&	7438.045254	&	7506.497884	&	10854	&	setup1	&	BHr	&	142990	&	44	&	8562.978073	&	8569.834368	&	3145	&	setup1	&	BLb	\\
109026	&	15	&	7438.166777	&	7450.557591	&	5377	&	setup1	&	BLb	&	142990	&	44	&	8569.894533	&	8596.684458	&	12271	&	setup2	&	BLb	\\
109026	&	15	&	7457.249672	&	7526.084660	&	6447	&	setup2	&	BLb	&	142990	&	44	&	8596.959920	&	8610.844410	&	6236	&	setup3	&	BLb	\\
109026	&	15	&	7526.283363	&	7549.254528	&	11201	&	setup3	&	BLb	&	142990	&	44	&	8610.905102	&	8669.198978	&	23466	&	setup4	&	BLb	\\
109026	&	15	&	7550.144184	&	7553.127357	&	1613	&	setup4	&	BLb	&	142990	&	44	&	8669.199718	&	8720.671688	&	15283	&	setup5	&	BLb	\\
109026	&	15	&	7553.187169	&	7555.132863	&	1173	&	setup5	&	BLb	&	142990	&	44	&	8722.114378	&	8724.335484	&	911	&	setup6	&	BLb	\\
109026	&	15	&	7555.192675	&	7585.493594	&	15523	&	setup6	&	BLb	&	150549	&	18	&	7604.212806	&	7610.360875	&	2592	&	setup2	&	BTr	\\
109026	&	15	&	7442.061824	&	7520.063959	&	22186	&	setup3	&	BTr	&	150549	&	18	&	7611.172854	&	7624.143975	&	4061	&	setup3	&	BTr	\\
109026	&	15	&	7520.124540	&	7592.481186	&	35796	&	setup4	&	BTr	&	150549	&	18	&	7624.205747	&	7646.116570	&	7805	&	setup4	&	BTr	\\
109026	&	62	&	9296.527872	&	9297.421978	&	429	&	setup1	&	BTr	&	152564	&	18	&	7611.172909	&	7624.144072	&	4179	&	setup3	&	BTr	\\
109026	&	62	&	9297.482989	&	9305.264907	&	3453	&	setup2	&	BTr	&	152564	&	18	&	7624.205845	&	7646.116727	&	8263	&	setup4	&	BTr	\\
109026	&	62	&	9305.534079	&	9449.949762	&	60329	&	setup3	&	BTr	&	155203	&	27	&	7837.302989	&	8013.134183	&	45900	&	setup1	&	BAb	\\
120640	&	35	&	8245.375517	&	8247.414308	&	985	&	setup4	&	BHr	&	155203	&	54	&	8928.391454	&	9114.175778	&	26842	&	setup1	&	BAb	\\
120640	&	35	&	8248.012313	&	8253.467189	&	1885	&	setup5	&	BHr	&	155203	&	63	&	9310.518552	&	9479.146895	&	21767	&	setup1	&	BAb	\\
122532	&	35	&	8236.416745	&	8241.008196	&	2108	&	setup2	&	BHr	&	157792	&	3	&	6776.590195	&	6806.444533	&	2960	&	setup1	&	UBr	\\
122532	&	35	&	8241.398726	&	8245.322090	&	2345	&	setup3	&	BHr	&	157792	&	27	&	7837.302679	&	8013.133384	&	44560	&	setup1	&	BAb	\\
122532	&	35	&	8245.375812	&	8247.414602	&	989	&	setup4	&	BHr	&	157792	&	27	&	7971.395833	&	7973.368883	&	1549	&	setup1	&	BHr	\\
122532	&	35	&	8248.012606	&	8253.467478	&	1897	&	setup5	&	BHr	&	157792	&	27	&	7973.499836	&	7982.439861	&	4967	&	setup2	&	BHr	\\
122532	&	35	&	8257.640968	&	8271.501581	&	5401	&	setup1	&	BTr	&	157792	&	37	&	8198.545399	&	8277.191879	&	13690	&	setup1	&	BAb	\\
125823	&	2	&	6756.591857	&	6887.908837	&	16721	&	setup4	&	BAb	&	157792	&	54	&	8928.391349	&	9114.175888	&	20615	&	setup1	&	BAb	\\
125823	&	2	&	6835.821147	&	6841.785639	&	4953	&	setup1	&	BTr	&	157792	&	63	&	9310.587335	&	9479.147007	&	16882	&	setup1	&	BAb	\\
125823	&	2	&	6748.466415	&	6887.470095	&	63298	&	setup4	&	UBr	&	157919	&	3	&	6776.590129	&	6806.442322	&	2805	&	setup1	&	UBr	\\
128898	&	2	&	6756.590088	&	6887.910122	&	39788	&	setup4	&	BAb	&	157919	&	27	&	7837.302853	&	8013.133440	&	43984	&	setup1	&	BAb	\\
128898	&	2	&	6820.812181	&	6847.459022	&	4003	&	setup1	&	BLb	&	157919	&	27	&	7971.395852	&	7973.368904	&	1553	&	setup1	&	BHr	\\
128898	&	2	&	6835.821002	&	6841.785675	&	4941	&	setup1	&	BTr	&	157919	&	27	&	7973.499857	&	7982.439892	&	4962	&	setup2	&	BHr	\\
128898	&	2	&	6742.188018	&	6887.471371	&	64187	&	setup4	&	UBr	&	157919	&	37	&	8198.545339	&	8277.190188	&	13419	&	setup1	&	BAb	\\
128898	&	15	&	7423.395470	&	7443.063159	&	1018	&	setup1	&	BAb	&	157919	&	54	&	8928.391526	&	9114.172419	&	13019	&	setup1	&	BAb	\\
128898	&	15	&	7443.336187	&	7448.226499	&	1273	&	setup2	&	BAb	&	157919	&	63	&	9310.587269	&	9479.838789	&	12459	&	setup1	&	BAb	\\
128898	&	15	&	7448.286819	&	7470.188121	&	5908	&	setup3	&	BAb	&	159876	&	27	&	7854.576248	&	7910.468510	&	5753	&	setup1	&	BLb	\\
128898	&	15	&	7470.248205	&	7504.286610	&	2770	&	setup4	&	BAb	&	162374	&	54	&	8982.500619	&	8988.436436	&	2658	&	setup4	&	BTr	\\
128898	&	15	&	7520.798614	&	7521.993325	&	326	&	setup5	&	BAb	&	165040	&	18	&	7604.213365	&	7610.361704	&	2620	&	setup2	&	BTr	\\
128898	&	15	&	7522.750872	&	7536.004061	&	2606	&	setup6	&	BAb	&	168733	&	54	&	8978.475015	&	8982.438689	&	1549	&	setup3	&	BTr	\\
128898	&	15	&	7438.166311	&	7450.557354	&	5139	&	setup1	&	BLb	&	169467	&	27	&	7837.299534	&	8013.135520	&	46416	&	setup1	&	BAb	\\
128898	&	15	&	7457.249564	&	7526.085668	&	6989	&	setup2	&	BLb	&	169467	&	27	&	7846.287481	&	7898.245144	&	7810	&	setup1	&	UBr	\\
128898	&	15	&	7526.284373	&	7549.255669	&	11194	&	setup3	&	BLb	&	169467	&	54	&	8928.390352	&	9114.177126	&	27667	&	setup1	&	BAb	\\
128898	&	15	&	7550.145327	&	7553.128503	&	1615	&	setup4	&	BLb	&	169467	&	54	&	8978.474788	&	8982.438450	&	1521	&	setup3	&	BTr	\\
128898	&	15	&	7553.188316	&	7555.134010	&	1177	&	setup5	&	BLb	&	169467	&	54	&	8988.497298	&	9056.717478	&	27395	&	setup5	&	BTr	\\
128898	&	15	&	7555.193823	&	7585.494601	&	15528	&	setup6	&	BLb	&	169467	&	54	&	9056.718188	&	9129.434250	&	31305	&	setup6	&	BTr	\\
128898	&	15	&	7428.682458	&	7429.642491	&	193	&	setup1	&	BTr	&	169467	&	63	&	9310.516779	&	9479.148713	&	23148	&	setup1	&	BAb	\\
128898	&	15	&	7430.118182	&	7441.996591	&	3929	&	setup2	&	BTr	&	173648	&	17	&	7633.990911	&	7636.072662	&	726	&	setup2	&	BLb	\\
128898	&	15	&	7442.061897	&	7520.064906	&	23138	&	setup3	&	BTr	&	173648	&	17	&	7636.272765	&	7645.478077	&	5272	&	setup3	&	BLb	\\
128898	&	15	&	7520.125018	&	7592.482121	&	40050	&	setup4	&	BTr	&	173648	&	17	&	7512.729188	&	7520.040519	&	3246	&	setup2	&	BTr	\\
128898	&	62	&	9296.528017	&	9297.422140	&	430	&	setup1	&	BTr	&	173648	&	17	&	7520.101574	&	7553.007719	&	12723	&	setup3	&	BTr	\\
128898	&	62	&	9297.483153	&	9305.265219	&	3463	&	setup2	&	BTr	&	173648	&	17	&	7553.069260	&	7651.966953	&	49108	&	setup4	&	BTr	\\
128898	&	62	&	9305.534396	&	9449.950711	&	60582	&	setup3	&	BTr	&	173648	&	17	&	7590.399609	&	7595.708940	&	1372	&	setup1	&	UBr	\\
135379	&	2	&	6756.590342	&	6888.405032	&	16596	&	setup4	&	BAb	&	173648	&	17	&	7595.771182	&	7644.873843	&	18051	&	setup2	&	UBr	\\
135379	&	2	&	6835.821506	&	6841.786163	&	4951	&	setup1	&	BTr	&	173648	&	38	&	8330.620481	&	8333.049515	&	1365	&	setup1	&	BLb	\\
135379	&	2	&	6742.188153	&	6887.471590	&	64375	&	setup4	&	UBr	&	173648	&	38	&	8333.524944	&	8354.900329	&	10296	&	setup2	&	BLb	\\
141556	&	9	&	7199.528397	&	7263.330883	&	20547	&	setup3	&	BHr	&	173648	&	38	&	8354.960489	&	8391.129011	&	20003	&	setup3	&	BLb	\\
141556	&	9	&	7100.407988	&	7103.446596	&	391	&	setup1	&	BLb	&	173648	&	38	&	8272.072606	&	8274.534419	&	105	&	setup1	&	BTr	\\
141556	&	9	&	7106.071765	&	7110.989586	&	1017	&	setup2	&	BLb	&	173648	&	38	&	8274.665940	&	8286.474143	&	5285	&	setup2	&	BTr	\\
141556	&	9	&	7113.897501	&	7140.328987	&	4947	&	setup3	&	BLb	&	173648	&	38	&	8286.672340	&	8304.558767	&	6418	&	setup3	&	BTr	\\
141556	&	9	&	7142.193637	&	7197.123402	&	9839	&	setup4	&	BLb	&	173648	&	38	&	8304.688381	&	8371.574753	&	26456	&	setup4	&	BTr	\\
141556	&	9	&	7197.189998	&	7218.161941	&	6222	&	setup5	&	BLb	&	174638	&	17	&	7633.991177	&	7636.072918	&	727	&	setup2	&	BLb	\\
141556	&	9	&	7218.564797	&	7261.182097	&	10472	&	setup6	&	BLb	&	174638	&	17	&	7636.273020	&	7645.478285	&	5275	&	setup3	&	BLb	\\
        \hline
    \end{tabular}
    \end{adjustbox} 
\end{table*}

\begin{table*}
    \centering
    \caption{Table \ref{obs_log} continued.}
    \begin{adjustbox}{max width=\textwidth}
    \begin{tabular}{cccccccccccccc}
    \hline
    HD & Field & HJD(start) & HJD(end) & $N_\mathrm{Obs}$ & Mode & Sat & HD & Field & HJD(start) & HJD(end) & $N_\mathrm{Obs}$ & Mode & Sat \\
        \hline
    174638	&	17	&	7512.730253	&	7520.040687	&	3261	&	setup2	&	BTr	&	201601	&	55	&	9012.016128	&	9052.335110	&	10995	&	setup1	&	BHr	\\
174638	&	17	&	7520.101742	&	7553.008041	&	12828	&	setup3	&	BTr	&	202444	&	4	&	6851.471247	&	6939.917590	&	20456	&	setup1	&	BLb	\\
174638	&	17	&	7553.069583	&	7651.967125	&	49206	&	setup4	&	BTr	&	202444	&	4	&	6948.567100	&	6986.485678	&	3879	&	setup2	&	BLb	\\
174638	&	17	&	7653.671500	&	7664.799931	&	619	&	setup5	&	BTr	&	202444	&	4	&	6844.690930	&	6924.453737	&	25827	&	setup1	&	BTr	\\
174638	&	17	&	7590.399989	&	7595.709317	&	1413	&	setup1	&	UBr	&	202444	&	4	&	6820.675129	&	6839.655773	&	2689	&	setup1	&	UBr	\\
174638	&	17	&	7595.771559	&	7644.874055	&	17924	&	setup2	&	UBr	&	202444	&	10	&	7189.406761	&	7258.809126	&	8684	&	setup4	&	BAb	\\
174638	&	38	&	8317.919734	&	8318.333772	&	246	&	setup1	&	BHr	&	202444	&	10	&	7261.813474	&	7266.172690	&	2149	&	setup1	&	BLb	\\
174638	&	38	&	8318.478755	&	8325.471560	&	1782	&	setup2	&	BHr	&	204188	&	55	&	9012.015343	&	9051.930278	&	10854	&	setup1	&	BHr	\\
174638	&	38	&	8326.075898	&	8415.335780	&	13978	&	setup3	&	BHr	&	205924	&	55	&	9012.015754	&	9052.335092	&	10990	&	setup1	&	BHr	\\
174638	&	38	&	8330.620853	&	8333.049883	&	1371	&	setup1	&	BLb	&	206155	&	55	&	9012.015569	&	9051.930733	&	10793	&	setup1	&	BHr	\\
174638	&	38	&	8333.525311	&	8354.900636	&	10304	&	setup2	&	BLb	&	209790	&	11	&	7263.540832	&	7265.298397	&	140	&	setup1	&	BHr	\\
174638	&	38	&	8354.960796	&	8391.129132	&	20016	&	setup3	&	BLb	&	209790	&	11	&	7266.363251	&	7288.358803	&	4412	&	setup2	&	BHr	\\
174638	&	38	&	8272.072884	&	8274.534708	&	111	&	setup1	&	BTr	&	209790	&	11	&	7289.023254	&	7313.311944	&	1200	&	setup3	&	BHr	\\
174638	&	38	&	8274.666229	&	8286.474476	&	5277	&	setup2	&	BTr	&	209790	&	11	&	7294.808988	&	7296.137226	&	1063	&	setup2	&	BLb	\\
174638	&	38	&	8286.672673	&	8304.559140	&	6433	&	setup3	&	BTr	&	211336	&	29	&	7936.630652	&	7940.942017	&	1196	&	setup1	&	BTr	\\
174638	&	38	&	8304.688755	&	8371.574984	&	25622	&	setup4	&	BTr	&	211336	&	29	&	7941.000913	&	7945.106495	&	1370	&	setup2	&	BTr	\\
174638	&	38	&	8371.704831	&	8410.601516	&	3714	&	setup5	&	BTr	&	211336	&	29	&	7945.164204	&	7951.931119	&	1869	&	setup3	&	BTr	\\
175362	&	54	&	8946.547249	&	8954.538014	&	3301	&	setup1	&	BTr	&	211336	&	29	&	7952.065952	&	7960.056408	&	3915	&	setup4	&	BTr	\\
175362	&	54	&	8954.598793	&	8978.415294	&	11018	&	setup2	&	BTr	&	211336	&	29	&	7960.113636	&	8023.516120	&	34793	&	setup5	&	BTr	\\
175362	&	54	&	8978.474479	&	8982.438182	&	1506	&	setup3	&	BTr	&	211336	&	29	&	8023.855050	&	8102.032805	&	22013	&	setup6	&	BTr	\\
175362	&	54	&	8982.499682	&	8988.435594	&	2655	&	setup4	&	BTr	&	211336	&	57	&	9105.600883	&	9125.474602	&	2441	&	setup1	&	BHr	\\
175362	&	54	&	8988.497095	&	9056.717986	&	27580	&	setup5	&	BTr	&	211336	&	57	&	9126.139264	&	9157.896438	&	4282	&	setup2	&	BHr	\\
175362	&	54	&	9056.718696	&	9129.439816	&	31466	&	setup6	&	BTr	&	223128	&	57	&	9126.138974	&	9157.896456	&	4180	&	setup2	&	BHr	\\
176723	&	54	&	8954.598668	&	8978.415179	&	11008	&	setup2	&	BTr	&	225289	&	11	&	7263.540218	&	7265.298526	&	130	&	setup1	&	BHr	\\
182255	&	17	&	7553.069856	&	7651.967929	&	48929	&	setup4	&	BTr	&	225289	&	11	&	7266.363403	&	7288.359400	&	4018	&	setup2	&	BHr	\\
182255	&	17	&	7653.672293	&	7664.800638	&	622	&	setup5	&	BTr	&														\\
182568	&	17	&	7512.728838	&	7520.040291	&	3254	&	setup2	&	BTr	&														\\
182568	&	17	&	7520.101346	&	7553.008051	&	12703	&	setup3	&	BTr	&														\\
182568	&	17	&	7553.069594	&	7651.967888	&	49134	&	setup4	&	BTr	&														\\
182568	&	17	&	7653.672260	&	7664.800656	&	629	&	setup5	&	BTr	&														\\
183056	&	38	&	8330.620905	&	8333.049968	&	1367	&	setup1	&	BLb	&														\\
183056	&	38	&	8333.525402	&	8354.901002	&	10293	&	setup2	&	BLb	&														\\
183056	&	64	&	9454.551204	&	9457.423672	&	1109	&	setup1	&	BTr	&														\\
183056	&	64	&	9457.485164	&	9548.503689	&	36617	&	setup2	&	BTr	&														\\
189178	&	10	&	7184.664391	&	7197.103573	&	10049	&	setup1	&	BTr	&														\\
189178	&	38	&	8330.620716	&	8333.049808	&	1330	&	setup1	&	BLb	&														\\
189178	&	38	&	8333.525248	&	8354.901112	&	10289	&	setup2	&	BLb	&														\\
189178	&	38	&	8354.961273	&	8391.130346	&	20015	&	setup3	&	BLb	&														\\
189178	&	38	&	8272.072965	&	8274.533400	&	120	&	setup1	&	BTr	&														\\
189178	&	38	&	8274.664923	&	8286.473350	&	5280	&	setup2	&	BTr	&														\\
189178	&	38	&	8286.671551	&	8304.558373	&	6475	&	setup3	&	BTr	&														\\
189849	&	10	&	7175.388625	&	7178.111475	&	221	&	setup1	&	BAb	&														\\
189849	&	10	&	7178.876771	&	7182.227824	&	581	&	setup2	&	BAb	&														\\
189849	&	10	&	7182.362608	&	7189.063804	&	883	&	setup3	&	BAb	&														\\
189849	&	10	&	7189.412398	&	7204.901093	&	979	&	setup5	&	BAb	&														\\
189849	&	10	&	7184.665261	&	7197.104547	&	10053	&	setup1	&	BTr	&														\\
189849	&	10	&	7185.502547	&	7203.508329	&	7430	&	setup2	&	UBr	&														\\
189849	&	17	&	7512.729088	&	7520.039654	&	3254	&	setup2	&	BTr	&														\\
198639	&	10	&	7175.386616	&	7178.111098	&	335	&	setup1	&	BAb	&														\\
198639	&	10	&	7178.875214	&	7182.226020	&	648	&	setup2	&	BAb	&														\\
198639	&	10	&	7182.360802	&	7189.061050	&	738	&	setup3	&	BAb	&														\\
198639	&	10	&	7189.410350	&	7204.899351	&	1377	&	setup5	&	BAb	&														\\
198639	&	10	&	7184.663450	&	7197.102755	&	8556	&	setup1	&	BTr	&														\\
198639	&	29	&	7936.632986	&	7940.945269	&	1577	&	setup1	&	BTr	&														\\
198639	&	29	&	7960.115222	&	8023.108455	&	36868	&	setup5	&	BTr	&														\\
198639	&	64	&	9454.551600	&	9457.424151	&	1107	&	setup1	&	BTr	&														\\
198639	&	64	&	9457.485645	&	9548.500172	&	36826	&	setup2	&	BTr	&														\\
201433	&	10	&	7273.776316	&	7286.929336	&	1741	&	setup2	&	BLb	&														\\
201433	&	10	&	7184.664044	&	7197.103540	&	9921	&	setup1	&	BTr	&														\\
201433	&	10	&	7197.161321	&	7232.998440	&	24170	&	setup2	&	BTr	&														\\
201433	&	10	&	7232.999145	&	7270.608675	&	25809	&	setup3	&	BTr	&														\\
201433	&	10	&	7270.609380	&	7306.513699	&	24272	&	setup4	&	BTr	&														\\
201433	&	10	&	7306.514405	&	7340.633799	&	21154	&	setup5	&	BTr	&														\\
        \hline
    \end{tabular}
    \end{adjustbox} 
\end{table*}

\section{Essential data for our sample stars}

Table \ref{table_master1} lists essential data for our sample stars. It is organised as follows:

\begin{itemize}
\item Column 1: HD number.
\item Column 2: Identifier from \citet{2009A&A...498..961R}.
\item Column 3: Right ascension (J2000). Positional information was taken from the $Gaia$ catalogue.
\item Column 4: Declination (J2000).
\item Column 5: Spectral classification from \citet{2009A&A...498..961R}.
\item Column 6: Mean $V$ magnitude.
\item Column 7: Mean $V$ magnitude error.
\item Column 8:Parallax.
\item Column 9: Parallax error.
\item Column 10: Absorption in the $V$ band.
\item Column 11: Mean effective temperature.
\item Column 12: Mean effective temperature error.
\item Column 13: Luminosity.
\item Column 14: Luminosity error.
\end{itemize}

\begin{table*}
\caption{Essential data for our sample stars, sorted by increasing right ascension. The columns denote: 
(1) HD number. (2) Identifier from \citet{2009A&A...498..961R}. (3) Right ascension (J2000; $Gaia$). (4) Declination (J2000; $Gaia$). 
(5) Spectral classification from \citet{2009A&A...498..961R}. (6) Mean $V$ magnitude. (7) Mean $V$ magnitude error.
(8) Parallax from \citet{2007A&A...474..653V} and \citet{2018A&A...616A...1G,2023A&A...674A...1G}. (9) Parallax error.
(10) Absorption in the $V$ band derived as described in Section \ref{section_astro}. 
(11) Mean effective temperature derived as described in Section \ref{section_astro}. (12) Mean effective temperature error.
(13) Luminosity derived as described in Section \ref{section_astro}. (14) Luminosity error.}  
\label{table_master1}
\begin{center}
\begin{adjustbox}{max width=\textwidth,angle=90}
\begin{tabular}{lllcclcccccccccccccc}
\hline
\hline
(1) & (2) & (3) & (4) & (5) & (6) & (7) & (8) & (9) & (10) & (11) & (12) & (13) & (14) \\
HD	&	ID\_RM09	&	RA(J2000) 	&	 Dec(J2000)    	&	SpT\_RM09	& $V$\,mag	&	e\_$V$\,mag	&
$\pi$ &	e\_$\pi$ & $A_V$	& $\log T_\mathrm{eff}$ &
e\_$\log T_\mathrm{eff}$ & $\log L/L_\odot$ & e\_$\log L/L_\odot$ \\
\hline
6961	&	1773	&	17.777378	&	+55.149816	&	A5-F0	&	4.328	&	0.001	&	23.88	&	0.11	&	0.003	&	8115	&	187	&	1.400	&	0.004	\\
11415	&	2870	&	28.599179	&	+63.670021	&	B3 He wk.	&	3.345	&	0.002	&	7.00	&	0.25	&	0.040	&	14391	&	542	&	3.293	&	0.032	\\
15633	&	3900	&	37.688126	&	+00.255530	&	A6-F2	&	5.998	&	0.004	&	11.59	&	0.05	&	0.034	&	8161	&	152	&	1.371	&	0.004	\\
18296	&	4560	&	44.322016	&	+31.934091	&	A0 Si Sr	&	5.102	&	0.004	&	9.34	&	0.13	&	0.090	&	10990	&	223	&	2.090	&	0.011	\\
18866	&	4730	&	44.699379	&	$-$64.071266	&	A4-A7	&	4.969	&	0.002	&	10.16	&	0.09	&	0.031	&	8362	&	294	&	1.900	&	0.008	\\
20320	&	5060	&	48.958428	&	$-$08.819531	&	A2-A9	&	4.801	&	0.002	&	27.40	&	0.14	&	0.006	&	7691	&	137	&	1.103	&	0.005	\\
22634	&	5770	&	53.603920	&	$-$65.764153	&	A2-	&	6.741	&	0.004	&	8.27	&	0.11	&	0.099	&	8311	&	166	&	1.395	&	0.012	\\
22920	&	5840	&	55.159723	&	$-$05.210733	&	B8 Si	&	5.529	&	0.014	&	5.24	&	0.09	&	0.031	&	13511	&	287	&	2.610	&	0.014	\\
23408	&	6000	&	56.456793	&	+24.367548	&	B7 He wk. Mn	&	3.869	&	0.003	&	7.67	&	0.31	&	0.156	&	12719	&	170	&	2.932	&	0.036	\\
23850	&	6100	&	57.290684	&	+24.053218	&	B8 He wk.	&	3.621	&	0.003	&	8.12	&	0.48	&	0.107	&	12155	&	212	&	2.918	&	0.047	\\
25823	&	6560	&	61.651822	&	+27.599677	&	B9 Sr Si	&	5.194	&	0.005	&	8.07	&	0.11	&	0.000	&	12654	&	290	&	2.273	&	0.018	\\
26961	&	6880	&	64.561204	&	+50.295260	&	A2 Si	&	4.592	&	0.003	&	8.55	&	0.23	&	0.000	&	9135	&	226	&	2.206	&	0.027	\\
27463	&	7050	&	64.087788	&	$-$60.948410	&	A0 Eu Cr	&	6.631	&	0.096	&	7.92	&	0.42	&	0.000	&	9331	&	673	&	1.474	&	0.060	\\
27628	&	7080	&	65.515164	&	+14.077106	&	A3-F2	&	5.710	&	0.004	&	22.03	&	0.07	&	0.046	&	7295	&	103	&	0.968	&	0.003	\\
27962	&	7160	&	66.372921	&	+17.927773	&	A1-A4	&	4.297	&	0.004	&	21.09	&	0.14	&	0.000	&	8513	&	580	&	1.522	&	0.006	\\
28319	&	7240	&	67.166065	&	+15.870787	&	A8 Sr	&	3.398	&	0.004	&	21.81	&	0.37	&	0.016	&	7975	&	223	&	1.859	&	0.017	\\
28527	&	7324	&	67.640606	&	+16.193904	&	A5-F0	&	4.763	&	0.003	&	22.19	&	0.14	&	0.033	&	8097	&	159	&	1.301	&	0.006	\\
29140	&	7450	&	68.913825	&	+10.160617	&	A3-A7	&	4.243	&	0.002	&	19.51	&	0.83	&	0.000	&	8014	&	159	&	1.624	&	0.042	\\
29305	&	7490	&	68.499106	&	$-$55.045029	&	A0 Si	&	3.262	&	0.004	&	19.34	&	0.31	&	0.034	&	11586	&	57	&	2.225	&	0.015	\\
29388	&	7520	&	69.539886	&	+12.510773	&	A3-	&	4.262	&	0.004	&	22.03	&	0.19	&	0.000	&	8238	&	288	&	1.498	&	0.009	\\
30780	&	7950	&	72.843972	&	+18.839704	&	A5-F0	&	5.085	&	0.003	&	26.12	&	0.75	&	0.025	&	7861	&	188	&	1.037	&	0.023	\\
32549	&	8280	&	76.142366	&	+15.403968	&	B9 Si Cr	&	4.680	&	0.010	&	7.81	&	0.14	&	0.000	&	9900	&	291	&	2.299	&	0.013	\\
33641	&	8570	&	78.357061	&	+38.484173	&	A3-F1	&	4.821	&	0.003	&	20.10	&	0.50	&	0.000	&	8013	&	57	&	1.354	&	0.025	\\
33904	&	8640	&	78.233121	&	$-$16.205533	&	B9 Hg Mn	&	3.279	&	0.002	&	19.18	&	0.33	&	0.005	&	11341	&	1744	&	2.191	&	0.015	\\
33959	&	8660	&	78.851513	&	+32.687653	&	A6-F0 dD	&	5.000	&	0.003	&	12.01	&	0.12	&	0.052	&	7812	&	226	&	1.752	&	0.009	\\
34452	&	8790	&	79.750189	&	+33.748270	&	B9 Si	&	5.374	&	0.004	&	7.05	&	0.11	&	0.000	&	12855	&	1574	&	2.345	&	0.013	\\
35039	&	8953	&	80.440614	&	$-$00.382460	&	B2 He	&	4.729	&	0.012	&	2.86	&	0.16	&	0.128	&	18941	&	508	&	3.885	&	0.048	\\
35497	&	9110	&	81.572974	&	+28.607533	&	B8 Cr Mn	&	1.650	&	0.008	&	24.36	&	0.34	&	0.000	&	12520	&	409	&	2.734	&	0.013	\\
36960	&	9780	&	83.761181	&	$-$06.002023	&	B0 Si	&	4.754	&	0.004	&	2.62	&	0.12	&	0.050	&	25434	&	2839	&	4.176	&	0.107	\\
40183	&	10720	&	89.881823	&	+44.947432	&	A1-A3	&	1.889	&	0.004	&	40.21	&	0.23	&	0.000	&	8943	&	317	&	1.941	&	0.007	\\
40292	&	10740	&	88.708590	&	$-$52.634431	&	A7-F0 dD	&	5.282	&	0.002	&	28.33	&	0.05	&	0.000	&	7290	&	162	&	0.904	&	0.001	\\
40312	&	10750	&	89.929319	&	+37.212920	&	A0 Si	&	2.645	&	0.029	&	17.18	&	0.41	&	0.000	&	10242	&	310	&	2.455	&	0.024	\\
53244	&	14650	&	105.939561	&	$-$15.633337	&	B8 Mn Hg	&	4.099	&	0.002	&	7.58	&	0.19	&	0.037	&	13398	&	1300	&	2.842	&	0.021	\\
54118	&	14860	&	106.076391	&	$-$56.749702	&	A0 Si	&	5.151	&	0.002	&	10.36	&	0.17	&	0.000	&	10215	&	468	&	1.886	&	0.015	\\
55719	&	15140	&	108.065758	&	$-$40.498857	&	A3 Sr Cr Eu	&	5.302	&	0.004	&	7.56	&	0.10	&	0.000	&	8804	&	269	&	2.020	&	0.009	\\
56022	&	15250	&	108.305474	&	$-$45.183114	&	A0 Si Cr Sr	&	4.857	&	0.074	&	17.36	&	0.08	&	0.000	&	9666	&	143	&	1.521	&	0.003	\\
56455	&	15400	&	108.691681	&	$-$46.849671	&	A0 Si	&	5.708	&	0.007	&	7.37	&	0.07	&	0.096	&	12583	&	135	&	2.194	&	0.008	\\
59635	&	16280	&	112.273620	&	$-$38.812002	&	B5 Si	&	5.396	&	0.003	&	5.08	&	0.09	&	0.000	&	13641	&	451	&	2.690	&	0.015	\\
64740	&	17750	&	118.265088	&	$-$49.612980	&	B2 He	&	4.626	&	0.006	&	4.08	&	0.11	&	0.037	&	22398	&	729	&	3.646	&	0.051	\\
67523	&	18660	&	121.885642	&	$-$24.304123	&	F2-F5 dD	&	2.816	&	0.002	&	51.40	&	0.24	&	0.000	&	6754	&	131	&	1.440	&	0.004	\\
73634	&	20470	&	129.410909	&	$-$42.989038	&	A7	&	4.131	&	0.002	&	1.52	&	0.18	&	0.019	&	8343	&	374	&	3.896	&	0.108	\\
74560	&	20840	&	130.605601	&	$-$53.113883	&	B3 Mg Si	&	4.822	&	0.004	&	6.65	&	0.09	&	0.015	&	14567	&	592	&	2.764	&	0.013	\\
81188	&	23050	&	140.528331	&	$-$55.010607	&	B3 He wk.	&	2.490	&	0.012	&	5.70	&	0.30	&	0.097	&	18192	&	416	&	4.120	&	0.046	\\
\hline
\end{tabular}  
\end{adjustbox}    
\end{center}                                                                                                                           \end{table*}

\begin{table*}
\caption{Table \ref{table_master1} continued .}  
\begin{center}
\begin{adjustbox}{max width=\textwidth,angle=90}
\begin{tabular}{lllcclcccccccccccccc}
\hline
\hline
(1) & (2) & (3) & (4) & (5) & (6) & (7) & (8) & (9) & (10) & (11) & (12) & (13) & (14) \\
HD	&	ID\_RM09	&	RA(J2000) 	&	 Dec(J2000)    	&	SpT\_RM09	& $V$\,mag	&	e\_$V$\,mag	&
$\pi$ &	e\_$\pi$ & $A_V$	& $\log T_\mathrm{eff}$ &
e\_$\log T_\mathrm{eff}$ & $\log L/L_\odot$ & e\_$\log L/L_\odot$ \\
\hline
82434	&	23440	&	142.674003	&	$-$40.466567	&	F1-	&	3.582	&	0.002	&	54.31	&	0.33	&	0.000	&	6933	&	98	&	1.058	&	0.007	\\
93030	&	26850	&	160.738990	&	$-$64.394407	&	B0 Si N P	&	2.730	&	0.002	&	7.16	&	0.21	&	0.098	&	24718	&	1023	&	4.116	&	0.098	\\
104671	&	30280	&	180.754787	&	$-$63.312902	&	A4-A9	&	4.317	&	0.002	&	14.59	&	0.13	&	0.282	&	7706	&	417	&	1.955	&	0.008	\\
109026	&	31640	&	188.115994	&	$-$72.133014	&	B5 He wk.	&	3.860	&	0.008	&	8.60	&	0.28	&	0.002	&	15272	&	172	&	2.981	&	0.027	\\
120640	&	34740	&	207.946624	&	$-$46.898735	&	B3 He	&	5.769	&	0.012	&	3.07	&	0.09	&	0.106	&	18350	&	444	&	3.352	&	0.028	\\
122532	&	35150	&	210.864253	&	$-$41.423416	&	B9 Si	&	6.085	&	0.004	&	6.54	&	0.12	&	0.000	&	11373	&	872	&	2.008	&	0.016	\\
125823	&	35910	&	215.759192	&	$-$39.511907	&	B5 He wk.	&	4.379	&	0.044	&	8.33	&	0.18	&	0.058	&	17911	&	441	&	2.996	&	0.020	\\
128898	&	36710	&	220.624786	&	$-$64.976145	&	A9 Sr Eu	&	3.177	&	0.010	&	60.99	&	0.23	&	0.077	&	7913	&	104	&	1.080	&	0.003	\\
135379	&	38420	&	229.377732	&	$-$58.801793	&	A3-	&	4.060	&	0.002	&	33.82	&	0.25	&	0.015	&	8938	&	223	&	1.227	&	0.007	\\
138769	&	39515	&	233.971742	&	$-$44.958478	&	B3 He wk	&	4.538	&	0.004	&	6.85	&	0.13	&	0.000	&	16704	&	316	&	3.001	&	0.017	\\
141556	&	40190	&	237.739690	&	$-$33.627320	&	B9 Y Hg	&	3.956	&	0.002	&	15.90	&	0.34	&	0.009	&	10608	&	1100	&	2.023	&	0.021	\\
142990	&	40530	&	239.645218	&	$-$24.831595	&	B6 He wk.	&	5.424	&	0.009	&	7.01	&	0.09	&	0.260	&	16816	&	367	&	2.736	&	0.010	\\
150549	&	42610	&	251.666465	&	$-$67.109749	&	A0 Si	&	5.122	&	0.004	&	4.96	&	0.10	&	0.085	&	12264	&	592	&	2.736	&	0.020	\\
152564	&	43220	&	254.891328	&	$-$69.268217	&	A0 Si	&	5.785	&	0.005	&	5.22	&	0.24	&	0.025	&	12160	&	571	&	2.400	&	0.034	\\
155203	&	43840	&	258.038437	&	$-$43.240430	&	A -F3	&	3.315	&	0.004	&	44.57	&	0.24	&	0.000	&	6665	&	32	&	1.379	&	0.004	\\
157792	&	44360	&	261.592560	&	$-$24.175813	&	A3-F0 Sr	&	4.156	&	0.003	&	40.03	&	0.27	&	0.040	&	7521	&	42	&	1.054	&	0.006	\\
157919	&	44410	&	261.838733	&	$-$29.867629	&	F3- dD	&	4.274	&	0.003	&	29.12	&	0.20	&	0.000	&	6785	&	71	&	1.344	&	0.006	\\
159876	&	45000	&	264.396477	&	$-$15.398810	&	A5-F1 dD?	&	3.528	&	0.003	&	32.88	&	0.26	&	0.061	&	7597	&	142	&	1.481	&	0.007	\\
162374	&	45840	&	268.056939	&	$-$34.799222	&	B7 He wk.	&	5.875	&	0.009	&	3.61	&	0.06	&	0.243	&	15578	&	761	&	3.039	&	0.019	\\
165040	&	46610	&	272.145251	&	$-$63.669384	&	A3-F1 Sr	&	4.322	&	0.003	&	24.36	&	0.13	&	0.063	&	7909	&	49	&	1.411	&	0.004	\\
168733	&	47280	&	275.721129	&	$-$36.669667	&	B8 Ti Sr	&	5.335	&	0.004	&	5.06	&	0.10	&	0.000	&	12037	&	845	&	2.582	&	0.017	\\
169467	&	47480	&	276.743295	&	$-$45.968688	&	B4 He	&	3.509	&	0.003	&	11.67	&	0.40	&	0.000	&	14623	&	604	&	2.793	&	0.028	\\
173648	&	48650	&	281.193293	&	+37.605207	&	A4-F1	&	4.328	&	0.002	&	20.66	&	0.13	&	0.125	&	8212	&	157	&	1.576	&	0.006	\\
174638	&	48890	&	282.519991	&	+33.362651	&	B8 He	&	3.607	&	0.009	&	3.60	&	0.18	&	0.630	&	15894	&	1107	&	4.125	&	0.042	\\
175362	&	49030	&	284.168739	&	$-$37.343361	&	B6 He wk. Si	&	5.373	&	0.005	&	6.84	&	0.19	&	0.085	&	16681	&	531	&	2.710	&	0.022	\\
176723	&	49330	&	285.823804	&	$-$38.253087	&	F1- Sr	&	5.717	&	0.004	&	15.30	&	0.06	&	0.079	&	7183	&	206	&	1.305	&	0.003	\\
182255	&	50370	&	290.712028	&	+26.262363	&	B7 He wk.	&	5.193	&	0.011	&	8.91	&	0.17	&	0.041	&	13820	&	113	&	2.313	&	0.018	\\
182568	&	50490	&	291.031637	&	+29.621384	&	B3 He wk.	&	4.981	&	0.008	&	3.62	&	0.14	&	0.302	&	17861	&	502	&	3.552	&	0.031	\\
183056	&	50610	&	291.538053	&	+36.317956	&	B9 Si	&	5.155	&	0.005	&	5.78	&	0.10	&	0.000	&	12022	&	331	&	2.530	&	0.016	\\
189178	&	52400	&	299.307820	&	+40.367827	&	B5 He wk.	&	5.455	&	0.015	&	2.83	&	0.07	&	0.277	&	14510	&	50	&	3.350	&	0.020	\\
189849	&	52650	&	300.275484	&	+27.753590	&	A5-A9	&	4.652	&	0.001	&	13.44	&	0.11	&	0.078	&	8112	&	203	&	1.801	&	0.007	\\
198639	&	55270	&	312.521287	&	+44.059872	&	A4-A7 dD	&	5.056	&	0.002	&	24.23	&	0.07	&	0.045	&	8055	&	116	&	1.115	&	0.003	\\
201433	&	56170	&	317.162163	&	+30.205547	&	B9 Si Mg	&	5.612	&	0.020	&	8.20	&	0.10	&	0.083	&	11321	&	213	&	2.024	&	0.009	\\
201601	&	56210	&	317.585615	&	+10.130926	&	A9 Sr Eu	&	4.697	&	0.025	&	28.24	&	0.15	&	0.036	&	7768	&	90	&	1.126	&	0.004	\\
202444	&	56410	&	318.698682	&	+38.047393	&	F0 Sr	&	3.732	&	0.007	&	49.58	&	0.46	&	0.000	&	6786	&	329	&	1.100	&	0.009	\\
204188	&	56880	&	321.611456	&	+19.375713	&	A5-F0	&	6.066	&	0.004	&	21.68	&	0.14	&	0.003	&	7714	&	80	&	0.797	&	0.006	\\
205924	&	57330	&	324.633576	&	+05.771870	&	A8	&	5.657	&	0.004	&	19.05	&	0.12	&	0.000	&	7436	&	165	&	1.085	&	0.006	\\
206155	&	57410	&	325.008013	&	+09.184813	&	A2-A5	&	6.978	&	0.012	&	7.75	&	0.09	&	0.000	&	8047	&	758	&	1.314	&	0.009	\\
209790	&	58390	&	330.949803	&	+64.628355	&	A3-F4	&	4.397	&	0.003	&	32.12	&	0.81	&	0.235	&	7469	&	133	&	1.226	&	0.018	\\
211336	&	58660	&	333.762505	&	+57.043795	&	A7-F1 dD Sr	&	4.177	&	0.002	&	38.16	&	0.24	&	0.000	&	7390	&	147	&	1.082	&	0.007	\\
223128	&	61164	&	356.653156	&	+66.782232	&	B3 He	&	5.938	&	0.003	&	2.53	&	0.04	&	0.619	&	20318	&	1402	&	3.769	&	0.015	\\
225289	&	61790	&	1.275763	&	+61.313979	&	B8 Hg Mn	&	5.790	&	0.003	&	4.73	&	0.05	&	0.166	&	12681	&	1300	&	2.575	&	0.009	\\
\hline
\end{tabular}  
\end{adjustbox}    
\end{center}                                                                                                                           \end{table*}


\bsp	
\label{lastpage}
\end{document}